\documentclass[12pt,english]{article}
\usepackage{times}
\usepackage[T1]{fontenc}
\usepackage{geometry}
\usepackage{graphicx}
\usepackage{setspace}
\usepackage{amssymb}

\makeatletter

\providecommand{\tabularnewline}{\\}

\usepackage{bm}

\usepackage{babel}
\makeatother
\begin{document}

\section*{\noindent Capacity-Driven Low-Interference Fast Beam Synthesis for
Next Generation Base Stations }

\noindent \vfill

\noindent G. Oliveri,$^{(1)}$ \emph{Senior Member, IEEE}, G. Gottardi,$^{(1)}$
\emph{Member, IEEE}, N. Anselmi,$^{(1)}$ \emph{Member, IEEE}, and
A. Massa,$^{(1)(2)(3)}$ \emph{Fellow, IEEE}

\noindent \vfill

\noindent {\footnotesize $^{(1)}$} \emph{\footnotesize CNIT} {\footnotesize -
\char`\"{}University of Trento\char`\"{} Research Unit}{\footnotesize \par}

\noindent {\footnotesize Via Sommarive 9, 38123 Trento - Italy}{\footnotesize \par}

\noindent \textit{\emph{\footnotesize E-mail:}} {\footnotesize \{}\emph{\footnotesize giacomo.oliveri}{\footnotesize ,}
\emph{\footnotesize giorgio.gottardi}{\footnotesize ,} \emph{\footnotesize nicola.anselmi}{\footnotesize ,}
\emph{\footnotesize andrea.massa}{\footnotesize \}@}\emph{\footnotesize unitn.it}{\footnotesize \par}

\noindent {\footnotesize Website:} \emph{\footnotesize www.eledia.org/eledia-unitn}{\footnotesize \par}

\noindent {\footnotesize ~}{\footnotesize \par}

\noindent {\footnotesize $^{(2)}$} \emph{\footnotesize ELEDIA Research
Center} {\footnotesize (}\emph{\footnotesize ELEDIA}{\footnotesize @}\emph{\footnotesize UESTC}
{\footnotesize - UESTC)}{\footnotesize \par}

\noindent {\footnotesize School of Electronic Engineering, Chengdu
611731 - China}{\footnotesize \par}

\noindent \textit{\emph{\footnotesize E-mail:}} \emph{\footnotesize andrea.massa@uestc.edu.cn}{\footnotesize \par}

\noindent {\footnotesize Website:} \emph{\footnotesize www.eledia.org/eledia}{\footnotesize -}\emph{\footnotesize uestc}{\footnotesize \par}

\noindent {\footnotesize ~}{\footnotesize \par}

\noindent {\footnotesize $^{(3)}$} \emph{\footnotesize ELEDIA Research
Center} {\footnotesize (}\emph{\footnotesize ELEDIA@TSINGHUA} {\footnotesize -
Tsinghua University)}{\footnotesize \par}

\noindent {\footnotesize 30 Shuangqing Rd, 100084 Haidian, Beijing
- China}{\footnotesize \par}

\noindent {\footnotesize E-mail:} \emph{\footnotesize andrea.massa@tsinghua.edu.cn}{\footnotesize \par}

\noindent {\footnotesize Website:} \emph{\footnotesize www.eledia.org/eledia-tsinghua}{\footnotesize \par}

\noindent \vfill

\newpage
\section*{Capacity-Driven Low-Interference Fast Beam Synthesis for Next Generation
Base Stations}

~

~

~

\begin{flushleft}G. Oliveri, G. Gottardi, N. Anselmi, and A. Massa\end{flushleft}

\vfill

\begin{abstract}
\noindent The problem of the real-time multiple-input multiple-output
(\emph{MIMO}) array control when requirements on capacity performance,
out-of-cell interference, and computational efficiency are simultaneously
enforced is addressed by means of an innovative hybrid beamforming
technique. The synthesis of the excitations of the \emph{MIMO} system
is first re-formulated as that of matching an ideal {}``hybrid''
pattern fitting \emph{capacity} or \emph{low-interference} constraints
along the angular coordinates. Then, a non-iterative processing scheme
is derived for each \emph{MIMO} beam where numerically-efficient synthesis
techniques are profitably combined. Representative results, from an
extensive numerical validation, are discussed to show, also comparatively,
the advantages and the current limitations of the proposed synthesis
method when dealing with different propagation scenarios, number of
transmitters/receivers, and noise levels.

\vfill
\end{abstract}
\noindent \textbf{Key words}: Phased arrays; MIMO Base Stations; 5G
Communications; Beampattern Synthesis; Zero-Forcing Processing.

\newpage
\section{Introduction and Rationale\label{sec:Introduction}}

\noindent Multiple-Input Multiple-Output (\emph{MIMO})-based antenna
arrays are expected to be a key component of future base stations
(\emph{BS}s) for 5G communications systems and beyond in order to
meet increasingly challenging requirements in terms of data rate and
reconfigurability \cite{Rappaport 2017}-\cite{Oliveri 2019}. The
emergence of \emph{MIMO} processing in antenna arrays will have a
significant impact on the conception and the implementation of the
design procedures for \emph{BS}s and their real-time control \cite{Hong 2017}\cite{Oliveri 2019}.
As a matter of fact, most beamforming methods and array synthesis
techniques have been traditionally conceived for the optimization
of free-space properties \cite{Mailloux 2005}\cite{Balanis 1997}
such as gain, sidelobe profile, power-transfer-efficiency \cite{Shinohara 2013}\cite{Oliveri 2013b},
and null positions, but there are other performance indexes to be
assessed when \emph{MIMO} processing is at hand \cite{Hong 2017}\cite{Oliveri 2019}.
Thus, an exhaustive revision of traditional array design techniques
as well as beamforming strategies is necessary to meet the needs and
the performance targets of next-generation \emph{BS}s \cite{Hong 2017}\cite{Oliveri 2019}.
In such a framework, innovative paradigms for capacity-oriented array
synthesis have been recently introduced \cite{Oliveri 2019}. By recasting
the design problem to the maximization of the link quality (i.e.,
the sum-rate capacity as a function of the \emph{BS} array, the user
terminal antenna features/locations, and the propagation environment),
a significant improvement of the communication performance with respect
to traditional array designs has been yielded \cite{Oliveri 2019}.
As a by-product, it has been also proved that the patterns synthesized
with usually-adopted methods (e.g., the popular zero forcing {[}\emph{ZF}{]}
technique \cite{Katlenberger 2008}\cite{Kyosti 2017}) generally
do not focus the power towards the receiver direction as in traditional
beamforming methods (e.g., Taylor, Dolph, Slepian synthesis techniques
\cite{Mailloux 2005}\cite{Balanis 1997}), but they rather exhibit
{}``distorted'' and multi-lobes shapes \cite{Oliveri 2019} even
when regular fully-populated architectures are used. This implies
that there is a fundamental challenge, from both the theoretical and
the practical viewpoint, to be addressed in future \emph{BS}s for
mobile cellular communications when profitably exploiting purely capacity-driven
beamforming methods. Indeed, an excellent information transfer rate
can be obtained at the expenses of a not \emph{a-priori} controllable
spatial distribution of the radiated power \cite{Oliveri 2019} that
may result in unacceptable sidelobes and, consequently, critical out-of-sector
interferences.

\noindent In principle, traditional beamforming techniques (e.g.,
the methods based on convex programming and/or evolutionary optimization
strategies \cite{Rocca 2016}-\cite{Jin 2007}) could be adapted to
jointly handle capacity-driven and out-of-cell interference requirements
in \emph{MIMO} \emph{BS}s. However, realistic massive \emph{MIMO}
scenarios require the synthesis and the real-time update of a large
number of beams with minimum efforts and maximum scalability. Consequently,
applying a local/global iterative optimization scheme for each beam
at every update step may be impractical and/or unfeasible due to the
associated computational costs \cite{Rocca 2009}\cite{Wolpert 1997}.

\noindent In this paper, an innovative scalable hybrid array synthesis
method is proposed, as a complement to existing techniques, to simultaneously
yield (\emph{i}) capacity-oriented radiation performance and (\emph{ii})
out-of-cell sidelobe control, while also featuring (\emph{iii}) a
low-cost processing. To address these contrasting objectives, the
integration of a capacity-driven weighting scheme and a low-sidelobe
constrained synthesis is proposed. More specifically, a computationally-efficient
non-iterative hybrid capacity/sidelobe-driven (\emph{{}``HCS}'')
approach is derived where, for each beam, (\emph{Step 1}) two auxiliary
sets of excitations and the associated patterns are firstly synthesized
according to the capacity-maximization strategy \cite{Oliveri 2019}
({}``Max $\mathcal{C}$'' strategy) and an interference-controlled
technique denoted as {}``Low $\mathcal{I}$'' method. Then (\emph{Step
2}), an ideal \emph{hybridized pattern} is defined by setting to the
{}``Max $\mathcal{C}$'' pattern and to the {}``Low $\mathcal{I}$''
pattern the in-sector and the out-of-sector angular portions of the
radiated power, respectively. Finally (\emph{Step 3}), the array excitations
are determined to radiate the best angular approximation of the ideal
\emph{hybridized pattern}.

\noindent It is worthwhile to notice that such an approach could be
potentially sub-optimal owing to the fact that no global optimization
is carried out despite the nonlinearity of the problem, but it is
a suitable and profitable solution for real systems and applications.
Indeed, the paradigm behind the \emph{HCS} method is motivated by
the following theoretical and practical argumentations: (\emph{i})
widely-adopted capacity-driven beamforming techniques (e.g., the \emph{ZF}
scheme \cite{Oliveri 2019}\cite{Katlenberger 2008}\cite{Kyosti 2017})
do not easily integrate sidelobe constraints to control inter-cell
interferences; (\emph{ii}) the iterative local/global optimization
of the capacity performance and of the sidelobe profiles for each
beam, even though ideally viable, may be practically unfeasible due
to the computational costs and the convergence issues when dealing
with real-time massive \emph{MIMO} operations; (\emph{iii}) a suitable
trade-off between capacity maximization, sidelobe control, and computational
efficiency is mandatory to derive a beamforming scheme of practical
interest; (\emph{v}) several non-iterative array design techniques
are available in the state-of-the-art literature to match the samples
of a reference ideal pattern \cite{Balanis 1997} and they can be
effectively customized to address the \emph{hybridized pattern} synthesis
of the \emph{Step 3} of the \emph{HCS}. Finally, (\emph{iv}) the \emph{HCS}
paradigm is fully general, so it can be seamlessly customized to accommodate
more/less computationally-demanding and performing techniques to solve
each step of its implementation according to the available computational
power, real-time needs, and expected link quality performance.

\noindent To the best of the authors' knowledge, the main innovative
contributions of this research work are (\emph{i}) the introduction
of a computationally-efficient excitation synthesis method to jointly
address the maximization of the capacity and the control of the sidelobes;
(\emph{ii}) the derivation of a real-time beamforming procedure easily
extendable to account for additional features beyond inter-cell interference
minimization in massive \emph{MIMO} processing scenarios (e.g., static/dynamic
jamming rejection, null positioning); (\emph{iii}) the numerical assessment
of the proposed approach in terms of link quality, rejection of out-of-sector
interferences, and computational complexity in comparison with widely-adopted
capacity-driven and sidelobe-minimizing methods, as well.

\noindent The outline of the paper is as follows. First, the beamforming
problem is formulated (Sect. \ref{sec:Problem-Formulation}), then
the hybrid synthesis technique is illustrated (Sect. \ref{sec:SbD-Method}).
A selected set of numerical examples, drawn from an extensive validation
study, is presented and discussed to give some insights to the interested
readers of the features and the potentialities of the proposed approach
also in comparison with both state-of-the-art traditional and capacity-oriented
techniques (Sect. \ref{sec:Numerical-Analysis-and}). Some conclusions
are finally drawn (Sect. \ref{sec:Conclusions-and-Remarks}).

\section{\noindent Problem Statement and Mathematical Formulation\label{sec:Problem-Formulation} }

\noindent With reference to the benchmark scenario in Fig. 1, let
us consider a transmitting linear array that operates as a \emph{BS}
antenna in a multi-user multi-antenna downlink communication mode%
\footnote{\noindent Such a benchmark architecture has been chosen according
to its relevance for 5G \emph{BS}s. However, the same paradigm can
be generalized to any arbitrary architecture.%
}. Under the assumptions of (\emph{i}) single-carrier band-pass digitally-modulated
signals at the transmitters observed for the duration of a single
pulse%
\footnote{\noindent The multi-carrier multi-pulse cases can be seamlessly derived
by taking into account the carrier/pulse indexes in each quantity
within the mathematical formulation.%
}, (\emph{ii}) equal power distribution among the downlink beams, and
(\emph{iii}) mutual incoherence among the signal and the noise for
each beam, the average \emph{MIMO} downlink sum-rate capacity $\mathcal{C}_{ave}$
of the \emph{BS} across $P$ propagation scenarios turns out to be
\cite{Oliveri 2019}\begin{equation}
\mathcal{C}_{ave}\triangleq\frac{1}{P}\sum_{p=1}^{P}\mathcal{C}^{\left(p\right)}\label{eq:average capacity}\end{equation}
where $\mathcal{C}^{\left(p\right)}$ is the capacity of the $p$-th
($p=1,...,P$) propagation scenario at the carrier frequency $f$
given by\begin{equation}
\mathcal{C}^{\left(p\right)}\triangleq\sum_{\chi=V,H}\sum_{r=1}^{R}\mathcal{C}_{\chi,r}^{\left(p\right)}\label{eq:capacity scenario}\end{equation}
$\mathcal{C}_{\chi,r}^{\left(p\right)}$ being the downlink capacity
of the $\chi$-th ($\chi\in\left\{ V,H\right\} $) polarization at
the $r$-th ($r=1,...,R$) receiving antenna equal to \cite{Oliveri 2019}\cite{Kyosti 2017}\begin{equation}
\mathcal{C}_{\chi,r}^{\left(p\right)}\triangleq\log_{2}\left(1+\frac{\left|\left[\mathbf{g}_{\chi,r}^{\left(p\right)}\right]^{'}\mathbf{w}_{\chi,r}^{\left(p\right)}\right|^{2}}{\mu_{\chi,r}^{\left(p\right)}+\frac{2R\sigma^{2}\left(f\right)}{\Omega}}\right)\label{eq:capacity beam}\end{equation}
where $\sigma^{2}\left(f\right)$ is the noise power, which is assumed
constant across all the receivers for the sake of simplicity, $\Omega$
is the total power radiated by the \emph{BS}, $\cdot^{'}$ stands
for the transpose operator, and $\mu_{\chi,r}^{\left(p\right)}$\begin{equation}
\mu_{\chi,r}^{\left(p\right)}\triangleq\left\{ \sum_{\psi=V,H}\sum_{q=1}^{R}\left|\left[\mathbf{g}_{\psi,q}^{\left(p\right)}\right]^{'}\mathbf{w}_{\psi,q}^{\left(p\right)}\right|^{2}\right\} -\left|\left(\mathbf{g}_{\chi,r}^{\left(p\right)}\right)^{T}\mathbf{w}_{\chi,r}^{\left(p\right)}\right|^{2}\label{eq:intra-cell interference}\end{equation}
 is the \emph{intra-cell multi-user interference} at the $r$-th ($r=1,...,R$)
receiving antenna on the $\chi$ ($\chi\in\left\{ V,H\right\} $)
polarization. Moreover, $\mathbf{g}_{\chi,r}^{\left(p\right)}\triangleq\left\{ g_{\chi,\psi}^{\left(p\right)}\left(\underline{\rho}_{t},\underline{\rho}_{r}\right);\, t=1,...,T;\,\psi\in\left\{ V,H\right\} \right\} $
($\chi\in\left\{ V,H\right\} $; $r=1,...,R$) is a frequency-domain
complex vector whose ($t$, $\psi$)-th entry, $g_{\chi,\psi}^{\left(p\right)}\left(\underline{\rho}_{t},\underline{\rho}_{r}\right)$,
is the Fourier transform at the frequency $f$ of the time-domain
Green's function modeling the electromagnetic propagation between
the $\psi$-polarization ($\psi\in\left\{ V,H\right\} $) of the $t$-th
($t=1,...,T$) transmitting antenna and the $\chi$-polarization ($\chi\in\left\{ V,H\right\} $)
of the $r$-th ($r=1,...,R$) receiving element in correspondence
with the $p$-th ($p=1,...,P$) scenario, $\underline{\rho}_{t}=\left(x_{t},y_{t},z_{t}\right)$
and $\underline{\rho}_{r}=\left(x_{r},y_{r},z_{r}\right)$ being the
location of the $t$-th ($t=1,...,T$) transmitting and the $r$-th
($r=1,...,R$) receiving antennas, respectively, while $\mathbf{w}_{\chi,r}^{\left(p\right)}$
$\triangleq$ \{$w_{\chi,\psi,r,t}^{\left(p\right)}$; $t=1,...,T$;
$\psi\in\left\{ V,H\right\} $\} ($\chi\in\left\{ V,H\right\} $;
$r=1,...,R$) is the excitation vector affording the $\left(\chi,r\right)$-beam
(i.e., the pattern responsible for the transfer of information towards
the $\chi$-polarization of the $r$-th receiving antenna), whose
($t$, $\psi$)-th entry, $w_{\chi,\psi,r,t}^{\left(p\right)}$, is
the Fourier transform of the linear excitation as applied to the $\psi$-polarization
($\psi\in\left\{ V,H\right\} $) of the $t$-th ($t=1,...,T$) transmitting
antenna to synthesize the $\left(\chi,r\right)$-beam ($\chi\in\left\{ V,H\right\} $;
$r=1,...,R$) at the frequency $f$ in the $p$-th ($p=1,...,P$)
scenario. 

\noindent It is worth remarking that (\ref{eq:average capacity})-(\ref{eq:intra-cell interference})
hold true regardless of the array architecture (e.g., regular, thinned,
sparse, clustered, overlapped) adopted in the \emph{BS} \cite{Rocca 2016}\cite{Oliveri 2019}.
On the other hand, while (\ref{eq:average capacity}) quantifies the
{}``quality'' of the connections between the \emph{BS} and the terminals
within the cell of coverage \cite{Kyosti 2017}\cite{Oliveri 2019},
it does not take into account the out-of-sector interferences, these
latter being a key issue in multi-cell scenarios. To give a more reliable/complete
figure of merit for the \emph{BS} performance, the term (\ref{eq:intra-cell interference})
may be generalized to account for the presence and the locations of
the users in the nearby cells, for instance, by extending the summation
also to the out-of-sector receivers. Unfortunately, the real-time
evaluation of such an index, which is needed for the pattern synthesis,
would require the \emph{BS} to continuously sense the channel to update
the Green's frequency-domain complex vector for all the out-of-cell
users, as well. This would considerably convolute the already demanding
massive \emph{MIMO} channel estimation process. Alternatively, the
out-of-sector total interference $\mathcal{I}_{ave}$ can be computed
as the average of all the out-of-cell interferences across the $P$
propagation scenarios\begin{equation}
\mathcal{I}_{ave}\triangleq\frac{1}{P}\sum_{p=1}^{P}\mathcal{I}^{\left(p\right)},\label{eq:average interference}\end{equation}
the $p$-th ($p=1,...,P$) contribution, $\mathcal{I}^{\left(p\right)}$,
being efficiently estimated as the normalized integral of the pattern
sidelobes outside the cell sector. More specifically,\begin{equation}
\mathcal{I}^{\left(p\right)}\triangleq\sum_{\chi=V,H}\sum_{r=1}^{R}\mathcal{I}_{\chi,r}^{\left(p\right)}\label{eq:out-of-cell interference}\end{equation}
where $\mathcal{I}_{\chi,r}^{\left(p\right)}\triangleq\frac{\int\int_{\left\{ \theta,\varphi\right\} \notin\Xi}\left|A_{\chi,r}^{\left(p\right)}\left(\theta,\varphi\right)\right|^{2}\mathrm{d}\theta\mathrm{d}\varphi}{\int\int_{\left\{ \theta,\varphi\right\} }\left|A_{\chi,r}^{\left(p\right)}\left(\theta,\varphi\right)\right|^{2}\mathrm{d}\theta\mathrm{d}\varphi}$
is the out-of-sector interference ratio of the $\left(\chi,r\right)$-beam
($\chi\in\left\{ V,H\right\} $; $r=1,...,R$) at the frequency $f$
in the $p$-th ($p=1,...,P$) scenario, $\Xi\triangleq\left\{ \theta_{\min}\leq\theta\leq\theta_{\max};\varphi_{\min}\leq\varphi\leq\varphi_{\max}\right\} $
is the angular sector defining the coverage cell, and $A_{\chi,r}^{\left(p\right)}\left(\theta,\varphi\right)$
is the the far-field pattern of the $\left(\chi,r\right)$-beam ($\chi\in\left\{ V,H\right\} $;
$r=1,...,R$) at the frequency $f$ in the $p$-th ($p=1,...,P$)
scenario given by\begin{equation}
A_{\chi,r}^{\left(p\right)}\left(\theta,\varphi\right)=\mathcal{F}\left[\mathbf{w}_{\chi,r}^{\left(p\right)}\right]\triangleq\sum_{\psi=V,H}\sum_{t=1}^{T}\left(w_{\chi,\psi,r,t}^{\left(p\right)}\times E_{\psi,t}\left(\theta,\varphi\right)\exp\left[j\frac{2\pi f}{c_{0}}\left(\underline{\rho}_{t}\cdot\widehat{\underline{\rho}}\right)\right]\right)\label{eq:power pattern}\end{equation}
 $c_{0}$ being the speed of light. Moreover, $E_{\psi,t}\left(\theta,\varphi\right)$
is the embedded element factor of the $\psi$-polarization ($\psi\in\left\{ V,H\right\} $)
of the $t$-th ($t=1,...,T$) transmitting antenna, $\widehat{\underline{\rho}}$
$\triangleq$ ($\sin\theta\cos\varphi$, $\sin\theta\sin\varphi$,
$\cos\theta$) is the unit vector in spherical coordinates, while
$\mathcal{F}\left[\cdot\right]$ stands for the pattern/excitation
transformation operator.

\noindent Accordingly the \emph{Capacity-driven and Out-of-sector
Interference Mitigation} (\emph{COIM}) synthesis problem can be stated
as that of {}``\emph{computing the BS array excitations}, $\mathbf{w}_{\chi,r}^{\left(p\right)}$
($\chi\in\left\{ V,H\right\} $; $r=1,...,R$; $p=1,...,P$) \emph{so
that} (\emph{i}) \emph{the average MIMO downlink sum-rate capacity},
$\mathcal{C}_{ave}$, \emph{is maximized and} (\emph{ii}) \emph{the
out-of-sector total interference}, $\mathcal{I}_{ave}$, \emph{is
mitigated across all} $P$ \emph{propagation scenarios}.''

\section{\noindent \emph{HCS} Array Synthesis \emph{}Method\label{sec:SbD-Method}}

\noindent To address the \emph{COIM} synthesis problem stated in Sect.
\ref{sec:Problem-Formulation}, traditional excitation design methods
(e.g., Taylor, Dolph, Slepian techniques \cite{Mailloux 2005}\cite{Balanis 1997})
are not suitable since they are formulated to fit free-space performance
requirements (e.g., gain, mainlobe width/shape, null positions, sidelobe
level, etc ...) \cite{Mailloux 2005}\cite{Balanis 1997}. Otherwise,
synthesis techniques devoted to the maximization of the average \emph{MIMO}
downlink sum-rate capacity {[}e.g., dirty-paper coding (\emph{DPC}){]}
do not control the out-of-sector total interference \cite{Oliveri 2019}\cite{Kyosti 2017}\cite{Katlenberger 2008}.
The same holds true for sub-optimal capacity-oriented techniques based
on \emph{ZF}, which are widely employed because of the superior numerical
efficiency and the easier implementation with respect to \emph{DPC}
\cite{Oliveri 2019}. In principle, an alternative solution might
be that of recasting the synthesis problem at hand to an optimization
one by first defining a single/multi-objective cost function, which
depends on $\mathcal{C}_{ave}$ and $\mathcal{I}_{ave}$, then adopting
a suitable global optimization strategy possibly featuring an evolutionary
approach \cite{Rocca 2009}\cite{Rocca 2011}\cite{Haupt 2007}. However,
this is not viable in realistic massive \emph{MIMO} scenarios because
of the infeasibility/costs of several (i.e., one per each $\left(\chi,r\right)$-beam
and per time-instant) iterative optimizations to adjust/modify the
\emph{MIMO} patterns in real time according to the user movements
and the scenario variability.

\noindent In this paper, the \emph{COIM} array design is formulated
as a piecewise pattern matching problem for every $p$-th ($p=1,...,P$)
propagation scenario where the \emph{BS} excitations are set to approximate
the maximum-capacity pattern {[}i.e., $A_{\chi,r}^{\left(p\right)}\left(\theta,\varphi\right)\approx A_{\chi,r}^{\left(p\right),\mathcal{C}}\left(\theta,\varphi\right)${]}
and the low-sidelobe pattern {[}$A_{\chi,r}^{\left(p\right)}\left(\theta,\varphi\right)\approx A_{\chi,r}^{\left(p\right),\mathcal{I}}\left(\theta,\varphi\right)${]}
in the in-sector ($\left\{ \theta,\varphi\right\} \in\Xi$) and in
the out-of-cell ($\left\{ \theta,\varphi\right\} \notin\Xi$) portion
of each received $\left(\chi,r\right)$-beam ($\chi\in\left\{ V,H\right\} $;
$r=1,...,R$), respectively. Accordingly, the arising \emph{HCS} synthesis
method is then implemented as a three-step process where, for each
$\left(\chi,r\right)$-beam ($\chi\in\left\{ V,H\right\} $, $r=1,...,R$)
and $p$-th scenario ($p=1,...,P$), (Step 1 \emph{- Auxiliary Pattern
Setup}) the auxiliary/reference patterns $A_{\chi,r}^{\left(p\right),\mathcal{C}}\left(\theta,\varphi\right)$
and $A_{\chi,r}^{\left(p\right),\mathcal{I}}\left(\theta,\varphi\right)$
are first synthesized, then (Step 2 - \emph{Hybrid Pattern Definition})
they are combined into a piece-wise hybridized one $\widetilde{A}_{\chi,r}^{\left(p\right)}\left(\theta,\varphi\right)$,
and finally (Step 3 \emph{- Trade-Off Excitation Synthesis}) the tradeoff
excitations $\mathbf{w}_{\chi,r}^{\left(p\right)}$ are computed so
that the radiated pattern $A_{\chi,r}^{\left(p\right)}\left(\theta,\varphi\right)$
satisfies the matching condition $A_{\chi,r}^{\left(p\right)}\left(\theta,\varphi\right)\approx\widetilde{A}_{\chi,r}^{\left(p\right)}\left(\theta,\varphi\right)$. 

\noindent As for the {}``\emph{Auxiliary Pattern Setup}'' (Step
1) and the synthesis of the reference maximum-capacity pattern, $A_{\chi,r}^{\left(p\right),\mathcal{C}}\left(\theta,\varphi\right)$,
several different and available state-of-the-art techniques could
be employed, but since the excitations $\mathbf{w}_{\chi,r}^{\left(p\right)}$
must be computed in real time for each $\left(\chi,r\right)$-beam
($\chi\in\left\{ V,H\right\} $; $r=1,...,R$) and for every Green's
function variation {[}i.e., $p$-th ($p=1,...,P$) propagation scenario{]},
the \emph{ZF} scheme is adopted \cite{Oliveri 2019}\cite{Katlenberger 2008}\cite{Kyosti 2017}.
As a matter of fact, although sub-optimal, it guarantees a computationally-inexpensive
maximization of $\mathcal{C}_{\chi,r}^{\left(p\right)}$ ($\chi\in\left\{ V,H\right\} $,
$r=1,...,R$) by enforcing $\mu_{\chi,r}^{\left(p\right)}=0$ in (\ref{eq:capacity beam})
(i.e., no \emph{intra-cell multi-user interference}) \cite{Katlenberger 2008}\cite{Kyosti 2017}.
More in detail, the overall excitation set $\mathcal{W}^{\left(p\right),\mathcal{C}}$
$\triangleq$ \{$\mathbf{w}_{\chi,r}^{\left(p\right),\mathcal{C}}$;
$\chi\in\left\{ V,H\right\} $, $r=1,...,R$\} is defined as follows\begin{equation}
\mathcal{W}^{\left(p\right),\mathcal{C}}=\left[\mathcal{G}^{\left(p\right)}\right]^{\dagger}\label{eq:ZF excitations}\end{equation}
where $\mathcal{G}^{\left(p\right)}\triangleq\left\{ \mathbf{g}_{\chi,r}^{\left(p\right)};\, r=1,...,R;\,\chi\in\left\{ V,H\right\} \right\} $
is the Green's frequency-domain complex matrix, and the apex $\cdot^{\dagger}$
stands for the Moore-Penrose pseudo-inverse operator (i.e., $\mathcal{G}^{\dagger}\triangleq\mathcal{G}^{*}\left[\mathcal{G}\mathcal{G}^{*}\right]^{-1}$,
$\cdot^{*}$ being the conjugate transpose operator). Successively,
the patterns set \{$A_{\chi,r}^{\left(p\right),\mathcal{C}}\left(\theta,\varphi\right)$;
$\chi\in\left\{ V,H\right\} $; $r=1,...,R$; $p=1,...,P$\} is derived
by applying the pattern/excitation transformation operator to the
\emph{ZF}-based excitations $A_{\chi,r}^{\left(p\right),\mathcal{C}}\left(\theta,\varphi\right)=\mathcal{F}\left[\mathbf{w}_{\chi,r}^{\left(p\right),\mathcal{C}}\right]$.

\noindent The setting of $A_{\chi,r}^{\left(p\right),\mathcal{I}}\left(\theta,\varphi\right)$
to mitigate the out-of-sector total interference can be also yielded
with different standard methods \cite{Mailloux 2005}. To keep low
the computational complexity, while maximizing the radiation efficiency
and decreasing the sidelobe envelope to indirectly mitigate $\mathcal{I}_{ave}$
\cite{Mailloux 2005}, an isophoric (\emph{ISO}) excitation synthesis
scheme \cite{Oliveri 2019} is adopted. Accordingly, the auxiliary
pattern $A_{\chi,r}^{\left(p\right),\mathcal{I}}\left(\theta,\varphi\right)$
($\chi\in\left\{ V,H\right\} $; $r=1,...,R;$ $p=1,...,P$) is set
to that radiated by the excitation vector $\mathbf{w}_{\chi,r}^{\left(p\right),\mathcal{I}}$
(i.e., $A_{\chi,r}^{\left(p\right),\mathcal{I}}\left(\theta,\varphi\right)=\mathcal{F}\left[\mathbf{w}_{\chi,r}^{\left(p\right),\mathcal{I}}\right]$)
whose ($t,\,\psi$)-th ($t=1,...,T;\,\psi\in\left\{ V,H\right\} $)
entry is equal to\begin{equation}
w_{\chi,\psi,r,t}^{\left(p\right),\mathcal{I}}=\exp\left[\Phi_{tr}\right]\label{eq:ULA}\end{equation}
$\Phi_{tr}$ being the phase term that enables the $\chi$-polarized
($\chi\in\left\{ V,H\right\} $) beam to be steered towards the $r$-th
($r=1,...,R$) receiver location $\underline{\rho}_{r}$ \cite{Oliveri 2019}
(i.e., $\Phi_{tr}\triangleq\frac{2\pi f}{c_{0}}\left(\frac{\underline{\rho}_{t}\cdot\left(\underline{\rho}_{r}-\underline{\rho}_{t}\right)}{\left|\underline{\rho}_{r}-\underline{\rho}_{t}\right|}\right)${]}.

\noindent It is worth pointing out that, since the \emph{HCS} approach
does not depend on the methodologies to compute $A_{\chi,r}^{\left(p\right),\mathcal{C}}\left(\theta,\varphi\right)$
and $A_{\chi,r}^{\left(p\right),\mathcal{I}}\left(\theta,\varphi\right)$,
but rather on these latter as reference for the next steps, more computationally-intensive
strategies can be seamlessly adopted instead of (\ref{eq:ZF excitations})
and (\ref{eq:ULA}) according to the available computational resources
and desired performance.

\noindent Analogously, the \emph{}process (Step 2) \emph{}to define
the hybrid pattern $\widetilde{A}_{\chi,r}^{\left(p\right)}\left(\theta,\varphi\right)$
($\chi\in\left\{ V,H\right\} $, $r=1,...,R$, $p=1,...,P$), starting
from the references $A_{\chi,r}^{\left(p\right),\mathcal{C}}\left(\theta,\varphi\right)$
and $A_{\chi,r}^{\left(p\right),\mathcal{I}}\left(\theta,\varphi\right)$
of Step 1, can \emph{}be implemented in different ways. Owing to the
guidelines discussed above, a very simple solution would be preferred,
thus a straightforward piecewise approximation is chosen\begin{equation}
\widetilde{A}_{\chi,r}^{\left(p\right)}\left(\theta,\varphi\right)\triangleq\left\{ \begin{array}{ll}
A_{\chi,r}^{\left(p\right),\mathcal{C}}\left(\theta,\varphi\right) & \left\{ \theta,\varphi\right\} \in\Xi\\
A_{\chi,r}^{\left(p\right),\mathcal{I}}\left(\theta,\varphi\right) & \left\{ \theta,\varphi\right\} \notin\Xi\end{array}\right..\label{eq:raccordo}\end{equation}
Since (\ref{eq:raccordo}) may generate a non-continuous target pattern,
the exact matching between $A_{\chi,r}^{\left(p\right)}\left(\theta,\varphi\right)$
and $\widetilde{A}_{\chi,r}^{\left(p\right)}\left(\theta,\varphi\right)$
is inherently prevented because of the finite number of the \emph{BS}
array elements. Therefore, the excitations synthesis in the Step 3
is then aimed at finding $\mathbf{w}_{\chi,r}^{\left(p\right)}$ so
that the afforded pattern $A_{\chi,r}^{\left(p\right)}\left(\theta,\varphi\right)$
is an approximation of the hybrid and ideal one, $A_{\chi,r}^{\left(p\right)}\left(\theta,\varphi\right)\approx\widetilde{A}_{\chi,r}^{\left(p\right)}\left(\theta,\varphi\right)$,
for each $\left(\chi,r\right)$-beam ($\chi\in\left\{ V,H\right\} $;
$r=1,...,R$) and $p$-th ($p=1,...,P$) scenario. Following this
guideline, a simple, elegant, and insightful approach based on the
Woodward-Lawson (\emph{WL}) beam shaping technique \cite{Balanis 1997}
is exploited since (\emph{i}) it only needs the samples of the desired
pattern at various discrete locations, (\emph{ii}) it is simple to
implement, (\emph{iii}) it is extremely computationally-efficient,
and (\emph{iv}) it does not imply any iterative processing unlike
most alternative techniques \cite{Balanis 1997}\cite{Rocca 2009}
including deterministic \cite{Orchard 1985} or stochastic \cite{Haupt 2007}
strategies.

\noindent Under the assumption of regularly-spaced antennas {[}e.g.,
$\underline{\rho}_{t}\triangleq\left(0,\,\left(t-\frac{T-1}{2}\right)\times d,\,0\right)$
($t=1,...,T$), $d$ being inter-element spacing - Fig. 1{]} and negligible
edge effects (i.e., $E_{\psi,t}\left(\theta,\varphi\right)\approx E_{\psi}\left(\theta,\varphi\right)$,
$\psi\in\left\{ V,H\right\} $), the final array excitations are obtained
as follows \cite{Balanis 1997}\begin{equation}
w_{\chi,\psi,r,t}^{\left(p\right)}=\sum_{q=1}^{T}\left\{ \frac{\widetilde{A}_{\chi,r}^{\left(p\right)}\left(\theta_{q},\varphi_{q}\right)}{E_{\psi}\left(\theta_{q},\varphi_{q}\right)}\exp\left[-j\frac{2\pi f}{c_{0}}\left(y_{t}\sin\theta_{q}\sin\varphi_{q}\right)\right]\right\} \label{eq:WL method}\end{equation}
($\chi,\psi\in\left\{ V,H\right\} $; $r=1,...,R$; $t=1,...,T$)
where ($\theta_{q}$, $\varphi_{q}$) are the $q$-th ($q=1,...,Q$)
angular coordinates of azimuth pattern samples. Thanks to (\ref{eq:WL method}),
while generally $A_{\chi,r}^{\left(p\right)}\left(\theta,\varphi\right)\approx\widetilde{A}_{\chi,r}^{\left(p\right)}\left(\theta,\varphi\right)$,
the matching is ideal at the $Q$ sampling angles \cite{Balanis 1997},
$A_{\chi,r}^{\left(p\right)}\left(\theta_{q},\varphi_{q}\right)=\widetilde{A}_{\chi,r}^{\left(p\right)}\left(\theta_{q},\varphi_{q}\right)$
($q=1,...,Q$).

\noindent In summary, the proposed beamforming scheme solves the \emph{COIM}
problem by combining (\ref{eq:ZF excitations}), (\ref{eq:ULA}),
(\ref{eq:raccordo}), and (\ref{eq:WL method}). Consequently, the
computational complexity of the \emph{HCS} method only slightly increases
that of the popular and efficient \emph{ZF} scheme \cite{Katlenberger 2008}\cite{Kyosti 2017}
since the additional computations are concerned with a closed-form
isophoric weighting (\ref{eq:ULA}), a piecewise pattern combination
(\ref{eq:raccordo}), and a closed-form \emph{WL} synthesis (\ref{eq:WL method}).
Moreover, it is worth pointing out that the \emph{HCS} method does
not feature any configuration parameter, thus potential calibration
and robustness issues are avoided by definition.

\section{\noindent Numerical Assessment\label{sec:Numerical-Analysis-and}}

\noindent This section is aimed at illustrating the application of
the \emph{HCS} array synthesis method, at providing suitable guidelines
for an effective use in practical scenarios, and at assessing the
achievable performance in terms of capacity maximization and inter-cell
interference mitigation also in comparison with state-of-the-art beamforming
techniques for massive \emph{MIMO} processing and canonical array
synthesis approaches \cite{Oliveri 2019}\cite{Balanis 1997}. Towards
this end, different propagation scenarios (Fig. 2), number of transmitting/receiving
antennas, and signal-to-noise ratio (\emph{SNR}) conditions have been
considered, while the Green's function for each scenario has been
computed with a geometry-based stochastic method that exploits an
approximated ray-tracing approach \cite{Kyosti 2017} implemented
in the \emph{QuaDRiGa} physical-channel simulator \emph{}\cite{Rappaport 2017}\cite{Jaeckel 2012}-\cite{Jaeckel 2017}.

\noindent The first numerical experiment deals with a urban micro-cellular
scenario (\emph{UMi}) where $R=16$ terminals, equipped with dual-polarization
patch antennas, are served by a \emph{BS} \emph{MIMO} array, consisting
of $T=16$ $d=\frac{\lambda}{2}$-spaced patches featuring $\pm45$
{[}deg{]} polarization, located $10$ {[}m{]} above the ground. Moreover,
the coverage cell extends from $\varphi_{\min}=-60$ {[}deg{]} up
to $\varphi_{\max}=+60$ {[}deg{]}. As in \cite{Kyosti 2017}\cite{Oliveri 2019},
the propagation environment has been modeled according to the so-called
\emph{Madrid grid} {[}{}``Scenario 1'' - Fig. 2(\emph{a}){]} and
a generalized International Mobile Telecommunications-Advanced (\emph{IMT-A})
setup (\emph{NLOS IMT-A UMi}) has been simulated at $f=2.0$ GHz with
\emph{QuaDRiGa} \cite{Jaeckel 2012}-\cite{Jaeckel 2017}, by setting
$P=100$ and $SNR=20$ {[}dB{]} ($SNR\triangleq\frac{\Omega}{\sigma^{2}}$).
In Fig. 3(\emph{a}), the plots of the average capacity across $P$
random scenarios, $\mathcal{C}_{\chi,r}$ ($\mathcal{C}_{\chi,r}\triangleq\frac{1}{P}\sum_{p=1}^{P}\mathcal{C}_{\chi,r}^{\left(p\right)}$)
of the $\left(\chi,r\right)$-beam ($\chi\in\left\{ V,H\right\} $;
$r=1,...,R$) yielded with the \emph{HCS} (\ref{eq:WL method}), the
\emph{ZF} (\ref{eq:ZF excitations}), and the \emph{ISO} (\ref{eq:ULA})
methods are reported. As theoretically expected \cite{Oliveri 2019},
the \emph{ZF} approach guarantees the maximum communication performance
regardless of the beam index and the polarization state ($\Delta\mathcal{C}_{\chi,r}^{ZF-ISO}\in\left[3.2\times10^{1},1.3\times10^{5}\right]$
being $\Delta\mathcal{C}_{\chi,r}^{ZF-ISO}\triangleq\frac{\mathcal{C}_{\chi,r}^{ZF}}{\mathcal{C}_{\chi,r}^{ISO}}$)
thanks to a constructive exploitation of the propagation scenario
to increase the end-to-end information transfer. On the other hand,
while the \emph{ZF} significantly outperforms the \emph{ISO} method
in terms of average capacity ($\mathcal{C}_{ave}^{ZF}\approx187.75$
{[}bps/Hz{]} vs. $\mathcal{C}_{ave}^{ISO}\approx1.55$ {[}bps/Hz{]}
- Tab. I), the performance of the \emph{HCS} turns out are very close
to the \emph{ZF} one ($\Delta\mathcal{C}_{\chi,r}^{HCS-ZF}\in\left[0.57,0.86\right]$)
and better than that from the \emph{ISO} approach ( $\Delta\mathcal{C}_{\chi,r}^{HCS-ISO}\in\left[2.2\times10^{1},1.0\times10^{5}\right]$).
This trend is also confirmed by the corresponding overall average
capacity indexes ($\Delta\mathcal{C}_{ave}^{ZF-HCS}\approx1.44$ vs.
$\Delta\mathcal{C}_{ave}^{ISO-HCS}\approx1.19\times10^{-2}$ - Tab.
I). But what about the inter-cell interference of the radiated pattern?
The results in Fig. 3(\emph{b}) indicate that the \emph{HCS} technique
improves the average out-of-sector interference $\mathcal{I}_{\chi,r}$
($\mathcal{I}_{\chi,r}\triangleq\frac{1}{P}\sum_{p=1}^{P}\mathcal{I}_{\chi,r}^{\left(p\right)}$)
of the \emph{ZF} approach (i.e., $\mathcal{I}_{\chi,r}^{ZF}\in\left[-16.3,-8.1\right]$
{[}dB{]} vs. $\mathcal{I}_{\chi,r}^{HCS}\in\left[-28.0,-15.9\right]$
{[}dB{]}) with values not far from those of the \emph{ISO} method
being $\mathcal{I}_{\chi,r}^{ISO}\in\left[-27.6,-25.4\right]$ {[}dB{]}.
Such an outcome is further pointed out by the per-beam interference
mitigation difference, $\Delta\mathcal{I}_{\chi,r}^{ZF-HCS}$ ($\Delta\mathcal{I}_{\chi,r}^{ZF-HCS}\triangleq\mathcal{I}_{\chi,r}^{ZF}-\mathcal{I}_{\chi,r}^{HCS}$),
since $\Delta\mathcal{I}_{\chi,r}^{ZF-HCS}\in\left[6.2,16.4\right]$
{[}dB{]}, and from the corresponding average interference indexes
in Tab. I. Vice-versa, the average per-beam directivity $\mathcal{D}_{\chi,r}$
($\mathcal{D}_{\chi,r}\triangleq\frac{1}{P}\sum_{p=1}^{P}\mathcal{D}_{\chi,r}^{\left(p\right)}$,
$\mathcal{D}_{\chi,r}^{\left(p\right)}$ being the directivity of
the $\left(\chi,r\right)$-beam ($\chi\in\left\{ V,H\right\} $, $r=1,...,R$)
at frequency $f$ in the $p$-th ($p=1,...,P$) scenario) as well
as the overall directivity ($\Delta\mathcal{D}_{\chi,r}^{ISO-HCS}\approx8.2$
{[}dB{]}, being $\Delta\mathcal{D}_{\chi,r}^{ISO-HCS}\triangleq\mathcal{D}_{ave}^{ISO}-\mathcal{D}_{ave}^{HCS}$
- Tab. I) of both the \emph{HCS} and \emph{}the \emph{ZF} approaches
is considerably lower than that of the \emph{ISO} method {[}Fig. 3(\emph{c}){]}
as theoretically known \cite{Oliveri 2019}. As a matter of fact,
the \emph{HCS} beams exhibit distorted shapes with multi-lobes within
the \emph{BS} angular sector, while the sidelobes are well-controlled
in the out-of-sector angular directions as pictorially shown in Fig.
6(\emph{c}) for a representative case ($r=19$). For completeness
and illustrative purposes, let us analyze the behaviour of the patterns
$A_{\chi,r}^{\left(p\right),\mathcal{C}}\left(\theta,\varphi\right)$,
$A_{\chi,r}^{\left(p\right),\mathcal{I}}\left(\theta,\varphi\right)$,
and $A_{\chi,r}^{\left(p\right)}\left(\theta,\varphi\right)$ along
the $\theta=90$ {[}deg{]}-cut (Fig. 5) radiated by the transmitting
\emph{BS} when serving the ($r=19$)-th receiver using the excitations
in Fig. 4. One can notice that the \emph{HCS} technique carefully
approximates the \emph{ZF} beam within the in-sector angular region,
while it matches the \emph{ISO} pattern outside to reduces the sidelobes
when $\varphi\leq\varphi_{\min}$ or $\varphi\geq\varphi_{\max}$
{[}Figs. 5(\emph{a})-5(\emph{b}){]}. Moreover, the $A_{\chi,r}^{\left(p\right)}\left(\theta_{q},\varphi_{q}\right)$
exactly passes through the matching samples $\widetilde{A}_{\chi,r}^{\left(p\right)}\left(\theta_{q},\varphi_{q}\right)$
{[}i.e., blue line vs. black dots in Figs. 5(\emph{a})-5(\emph{b}){]}
as required by the \emph{WL} formulation.

\noindent The same conclusions are expected regardless of the users
location and their distributions. For a check, the second experiment
is concerned with the same \emph{BS} arrangement of the first example,
but assuming the propagation setup shown in Fig. 2(\emph{b}) ({}``\emph{Scenario
2}'') instead of that modeled by the {}``\emph{Scenario 1}'' {[}Fig.
2(\emph{a}){]}. The plots of the average capacity $\mathcal{C}_{\chi,r}$
{[}Fig. 7(\emph{a}){]} and of the average out-of-sector interference
$\mathcal{I}_{\chi,r}$ {[}Fig. 7(\emph{b}){]} confirm that the \emph{HCS}
method synthesizes radiation patterns featuring an almost optimal
information transfer {[}i.e., $\mathcal{C}_{\chi,r}^{ZF}\approx\mathcal{C}_{\chi,r}^{HCS}$
- Fig. 7(\emph{a}){]} and reduced inter-sector interferences {[}i.e.,
$\mathcal{I}_{\chi,r}^{ZF}\approx\mathcal{I}_{\chi,r}^{HCS}$ - Fig.
7(\emph{a}){]} with respect to the \emph{ZF} approach, while it outperforms
the \emph{ISO} scheme in terms of per-beam and average capacity {[}i.e.,
$\mathcal{C}_{\chi,r}^{ISO}<\mathcal{C}_{\chi,r}^{HCS}$ - Fig. 7(\emph{a}){]},
as confirmed by the values of the overall performance indexes ($\Delta\mathcal{C}_{ave}^{ZF-HCS}\approx2.2$
vs. $\Delta\mathcal{C}_{ave}^{ISO-HCS}\approx1.8\times10^{-2}$; $\Delta\mathcal{I}_{ave}^{ZF-HCS}\approx10$
{[}dB{]} - Tab. II), as well. On the other hand and once again, the
\emph{HCS} array yields values of $\mathcal{D}_{\chi,r}$ smaller
than those from the \emph{ISO} one {[}Fig. 7(\emph{c}){]} because
of multi-lobes pattern within the in-sector region ($\varphi_{\min}\le\varphi\le\varphi_{\max}$)
as visualized by the plots of both the 2D-cuts {[}Figs. 8(\emph{e})-Fig.
8(\emph{f}){]} and the 3D distributions {[}Fig. 9(\emph{c}) vs. Figs.
9(\emph{a})-9(\emph{b}){]} of the normalized power patterns $A_{\chi,r}^{\left(p\right),\mathcal{C}}\left(\theta,\varphi\right)$,
$A_{\chi,r}^{\left(p\right),\mathcal{I}}\left(\theta,\varphi\right)$,
and $A_{\chi,r}^{\left(p\right)}\left(\theta,\varphi\right)$ afforded
by the excitations reported in Figs. 8(\emph{a})-8(\emph{d}) and related
to the $r=1$ beam.

\noindent The next numerical experiment is aimed at assessing the
effectiveness of the \emph{HCS} paradigm as well as the dependence
of its performance on the number of transmitters, $T$, in the \emph{BS}.
With reference to the {}``\emph{Scenario 1}'' {[}Fig. 2(\emph{a}){]},
Figure 10 shows the behavior of $\mathcal{C}_{ave}$ {[}Fig. 10(\emph{a}){]},
$\mathcal{I}_{ave}$ {[}Fig. 10(\emph{b}){]}, and $\mathcal{D}_{ave}$
versus $T$. As it can be inferred from the plots, it turns out that
(\emph{i}) the interference mitigation capabilities of all techniques
increase when the array enlarges {[}Fig. 10(\emph{b}){]}; (\emph{ii})
the \emph{HCS} technique achieves capacity values very close to those
with the \emph{ZF} scheme regardless of the \emph{BS} size {[}i.e.,
$\Delta\mathcal{C}_{ave}^{ZF-HCS}\approx1.3$ - Fig. 10(\emph{a}){]},
but it improves the \emph{ZF} interference-mitigation features of
about $10$ {[}dB{]} {[}Fig. 10(\emph{b}){]}; (\emph{iii}) the \emph{HCS}
synthesis effectively and profitably exploits the additional degrees-of-freedom
of wider apertures by improving both the downlink capacity and the
interference suppression {[}Fig. 10(\emph{a}) and Fig. 10(\emph{b}){]};
(\emph{iv}) the free-space directivity of capacity-oriented strategies
{[}Fig. 10(\emph{c}){]} does not significantly improves when using
larger apertures since focusing the beam towards the receiver is known
to be a strongly sub-optimal capacity-maximization strategy when realistic/multi-path
propagation scenarios are at hand \cite{Oliveri 2019}.

\noindent The dependence of the performance indexes on the number
of receiving terminals $R$ has been analyzed next by keeping the
scenario in Fig. 2(\emph{a}) and setting $T=72$. In Fig. 11, the
plots of $\mathcal{C}_{ave}$, $\mathcal{I}_{ave}$, and $\mathcal{D}_{ave}$
as a function of $R$ are reported. As already proven in \cite{Oliveri 2019},
the \emph{ZF} sum-rate capacity does not monotonically enhances with
$R$ because of the unavoidable growth of the intra-cell multi-user
interference $\mu_{\chi,r}^{\left(p\right)}$ when the user density
increases, which causes a reduction of the propagation diversity of
the associated Green's functions {[}Fig. 11(\emph{a}){]}. As expected,
a similar trend arises when applying the \emph{HCS} {[}i.e., $\Delta\mathcal{C}_{ave}^{ZF-HCS}\in\left[1.25,4.5\right]$
- Fig. 11(\emph{a}){]}. The dependence of $\mathcal{I}_{ave}$ on
the number of receivers is also similar between \emph{HCS} and \emph{ZF}
methods, even though the former/hybrid one still strongly outperforms
the other {[}$\Delta\mathcal{\mathcal{I}}_{ave}^{ZF-HCS}\in\left[8,10\right]$
{[}dB{]} - Fig. 11(\emph{b}){]}. As for the comparison between the
\emph{HCS} and the \emph{ISO} performance, the same outcomes from
the analysis versus $T$ can be drawn {[}Fig. 11(\emph{c}){]}.

\noindent The final numerical test is aimed at evaluating the proposed
synthesis scheme when dealing with different propagation models and
$SNR$ values. More specifically, the \emph{QuaDRiGa} SW tool \cite{Jaeckel 2012}\cite{Jaeckel 2014}\cite{Jaeckel 2017}
has been configured to numerically compute the Green's functions of
a rural scenario with/without line-of-sight propagation conditions
according to the \emph{LOS 3GPP RMa} and the \emph{NLOS 3GPP RMa}
models \cite{3GPP 2017}, respectively. Accordingly, $\mathcal{C}_{ave}$
and $\mathcal{I}_{ave}$ have been computed in the benchmark $T=32$,
$R=16$ setup by varying the \emph{SNR} within the range $SNR\in\left[10,40\right]$
{[}dB{]} (Fig. 12). For comparison purposes, the values for the \emph{NLOS
IMT-A UMi} scenario are reported, as well. Notwithstanding the propagation
scenario and the noisy conditions, the \emph{HCS} further confirms
its effectiveness as optimal trade-off in terms of average downlink
capacity {[}$\mathcal{C}_{ave}$ - Fig. 12(\emph{a}){]} and out-of-sector
interference ratio {[}$\mathcal{I}_{ave}$ - Fig. 12(\emph{b}){]}
with respect to the other approaches. Moreover, it is worth noticing
that, even in rural scenarios and partially line-of-sight propagation
(i.e., \emph{LOS 3GPP RMa}), the capacity values yielded by the \emph{ISO}
technique are considerably smaller than those of the \emph{HCS} method
{[}Fig. 12(\emph{a}){]}. Such a result remarks that, as expected \cite{Oliveri 2019},
purely free-space oriented beamforming techniques such as (\ref{eq:ULA})
does not optimally transfer information also when low-scattering environments
with line-of-sight conditions are at hand {[}Fig. 12(\emph{a}){]}.

\noindent Of course, besides the trade-off optimality, the simplicity
and the computational efficiency are expected to be the other key
features of the proposed synthesis scheme. For assessment purposes,
the average overall synthesis time $\tau$ {[}$\tau\triangleq\frac{1}{P}\sum_{p=1}^{P}\tau^{\left(p\right)}$,
$\tau^{\left(p\right)}$ being the time spent for synthesizing the
$\left(\chi,r\right)$-beam ($\chi\in\left\{ V,H\right\} $; $r=1,...,R$)
excitations in the $p$-th ($p=1,...,P$) scenario: $\tau^{\left(p\right)}=\sum_{\chi=V,H}\sum_{r=1}^{R}\tau_{\chi,r}^{\left(p\right)}${]}
required by the \emph{ZF}, the \emph{ISO}, and the \emph{HCS} methods
when dealing with different setups of transmitters, $T$, and receivers,
$R$, has been reported in Fig. 13. For the sake of fairness, all
the \emph{CPU}-times refer to non-optimized Matlab implementations
executed on a single-core laptop PC with $1.60$ GHz-running \emph{CPU}.

\noindent As a representative test case, first the case of $R=16$
receivers located as in Fig. 2(\emph{a}) has been analyzed. Figure
13(\emph{a}) shows that, whatever the synthesis method, the synthesis
time $\tau$ slightly increases as the size of the transmitting antenna
widens (e.g., $\delta\tau^{ISO}\approx1.02$, $\delta\tau^{ZF}\approx1.05$,
and $\delta\tau^{HCS}\approx1.04$ being $\delta\tau\triangleq\frac{\left.\tau\right\rfloor _{T=104}}{\left.\tau\right\rfloor _{T=32}}$).
More important, the synthesis time for the \emph{HCS} is close to
that of the \emph{ZF} scheme (e.g., $\Delta\tau^{HCS-ZF}\approx1.19$
being $\Delta\tau^{HCS-ZF}\triangleq\frac{\left.\tau^{HCS}\right\rfloor _{T=104}}{\left.\tau^{ZF}\right\rfloor _{T=104}}$),
which is widely recognized as the most numerically efficient and easy
to implement approach \cite{Oliveri 2019}\cite{Katlenberger 2008}\cite{Kyosti 2017}.
Such a result is not surprising since it has been actually anticipated
in Sect. \ref{sec:SbD-Method} from the theoretical perspective. Indeed,
the \emph{HCS} implementation needs only closed-form steps (i.e.,
an isophoric weighting, a pattern combination, and a \emph{WL} synthesis)
besides \emph{ZF} one. For completeness, the results from the numerical
test with $T=72$ transmitters and varying the number of receivers,
$R$, are given in Fig. 13(\emph{b}). Despite a different dependance
of the \emph{CPU}-time on the number of terminals $R$ with respect
to the number of transmitting elements $T$, it is confirmed that
the \emph{HCS} implementation is simple and scalable with a computational
cost comparable to that of the popular \emph{ZF} paradigm, therefore
a suitable solution when real-time processing is required.

\section{\noindent Conclusions\label{sec:Conclusions-and-Remarks}}

\noindent An innovative pattern synthesis technique has been proposed
for the real-time \emph{MIMO} array control when capacity-oriented
radiation performance, out-of-cell sidelobe reduction, and low-complexity
processing are simultaneously required. Such an approach is based
on a non-iterative hybrid scheme where, for each beam, two auxiliary
sets of excitations and the associated patterns are firstly synthesized,
then an ideal hybrid pattern \emph{}is derived from their angular
combination so that the final array excitations are determined to
radiate the best angular approximation of this latter one. A set of
representative results, from a wide numerical analysis, has been presented
to discuss on the features and the potentialities of the proposed
\emph{HCS} synthesis as well as to assess its performance in terms
of downlink capacity, interference control, and computational efficiency
in comparison with other state-of-the-art popular beamforming methods,
as well.

\noindent To the best of the authors' knowledge, the main methodological
advancements of the paper with respect to the reference literature
on the topic include (\emph{i}) the introduction of a computationally-efficient
excitation synthesis method suitable for jointly handling the capacity-maximization
and the out-of-sector interference mitigation; (\emph{ii}) the derivation
of a real-time beamforming procedure that, thanks to its modularity,
can be easily extended to exploit additional computational resources
and/or more advanced/expensive single-step methods as well as to account
for additional requirements arising in next-generation massive \emph{MIMO}
scenarios (e.g., static/dynamic jamming rejection, null positioning);
(\emph{iii}) the numerical assessment of the proposed approach in
comparison with widely-adopted capacity-driven and sidelobe-minimizing
methods, as well, as a mandatory step for deriving suitable guidelines
towards a practical implementation of the \emph{HCS} strategy.

\noindent The numerical validation has shown that

\begin{itemize}
\item \noindent the use of the \emph{HCS} method implies a negligible capacity
reduction with respect to the popular \emph{ZF} algorithm regardless
of the propagation scenario, number of transmitters and receivers,
and noise level, while always outperforming the \emph{ISO} performance;
\item \noindent the \emph{HCS} considerably reduces the out-of-sector interferences
occurring when adopting the \emph{ZF} approach {[}Fig. 2(\emph{b}){]}
thanks to an effective shaping/control of the out-of-sector secondary
lobes;
\item the \emph{HCS} does not need calibration, notwithstanding the change
of the propagation scenario or noisy conditions, owing to the absence
of control parameters;
\item the computational complexity of the \emph{HCS} is similar to that
of the \emph{ZF} scheme, thus suitable for real-time operative conditions.
\end{itemize}
\noindent Future works, beyond the scope of the present paper, will
be aimed at generalizing the \emph{HCS} scheme to different array
architectures (e.g., sparse, thinned, clustered, overlapped, etc.)
and geometries (e.g., planar, conformal). Customizations to deal with
additional requirements and applicative guidelines in terms of both
intra- and inter-cell performance are also envisaged.

\section*{\noindent Acknowledgements}

\noindent This work benefited from the networking activities carried
out within the Project {}``Cloaking Metasurfaces for a New Generation
of Intelligent Antenna Systems (MANTLES)'' (Grant No. 2017BHFZKH)
funded by the Italian Ministry of Education, University, and Research
under the PRIN2017 Program (CUP: E64I19000560001), and within the
Project {}``SPEED'' (Grant No. 61721001) funded by the National
Science Foundation of China under the ChangJiang Visiting Professorship
Program. A. Massa wishes to thank E. Vico for her never-ending inspiration,
support, guidance, and help.

\newpage
\section*{FIGURE CAPTIONS}

\begin{itemize}
\item \textbf{Figure 1.} \emph{Problem Geometry.} Multi-user multi-antenna
downlink communication.
\item \textbf{Figure 2.} \emph{Benchmark Environments} ({}``Madrid Grid''
urban environment, \emph{NLOS IMT-A UMi} model). Location of the \emph{BS}
antenna (i.e., a transmitting array of \emph{S} radiating elements)
and of the \emph{L} terminals in (\emph{a}) the {}``\emph{Scenario
1}'' and (\emph{b}) the {}``\emph{Scenario 2}''.
\item \textbf{Figure 3.} \emph{Numerical Assessment} (\emph{NLOS IMT-A UMi}
model - \emph{Scenario 1}, $T=32$, $R=16$, $SNR=20$ {[}dB{]}, $P=100$).
Behaviour of \emph{}(\emph{a}) $\mathcal{C}_{\chi,r}$, (\emph{b})
$\mathcal{I}_{\chi,r}$, and (\emph{c}) $\mathcal{D}_{\chi,r}$ versus
the $r$-th ($r=1,...,R$) receiver index ($\chi\in\left\{ V,H\right\} $).
\item \textbf{Figure 4.} \emph{Numerical Assessment} (\emph{NLOS IMT-A UMi}
model - \emph{Scenario 1}, $T=32$, $R=16$, $SNR=20$ {[}dB{]}, $r=19$,
$p=1$). Plots of (\emph{a})(\emph{b}) the magnitude and (\emph{c})(\emph{d})
the phase of $w_{\chi,\psi,r,t}^{\left(p\right)}$ versus the $t$-th
($t=1,...,T$) array element index when (\emph{a})(\emph{c}) $\chi=\psi=V$
and (\emph{b})(\emph{d}) $\chi=\psi=H$. 
\item \textbf{Figure 5.} \emph{Numerical Assessment} (\emph{NLOS IMT-A UMi}
model - \emph{Scenario 1}, $T=32$, $R=16$, $SNR=20$ {[}dB{]}, $r=19$,
$p=1$). Normalized magnitude of $A_{\chi,r}^{\left(p\right),\mathcal{C}}\left(\theta,\varphi\right)$,
$A_{\chi,r}^{\left(p\right),\mathcal{I}}\left(\theta,\varphi\right)$,
and $A_{\chi,r}^{\left(p\right)}\left(\theta,\varphi\right)$ along
the $\theta=90$ {[}deg{]}-cut when (\emph{a}) $\chi=\psi=V$ and
(\emph{b}) $\chi=\psi=H$.
\item \textbf{Figure 6.} \emph{Numerical Assessment} (\emph{NLOS IMT-A UMi}
model - \emph{Scenario 1}, $T=32$, $R=16$, $SNR=20$ {[}dB{]}, $r=19$,
$p=1$, $\chi=\psi=V$). Normalized 3D magnitude of (\emph{a}) $A_{\chi,r}^{\left(p\right),\mathcal{C}}\left(\theta,\varphi\right)$,
(\emph{b}) $A_{\chi,r}^{\left(p\right),\mathcal{I}}\left(\theta,\varphi\right)$,
and (\emph{c}) $A_{\chi,r}^{\left(p\right)}\left(\theta,\varphi\right)$.
\item \textbf{Figure 7.} \emph{Numerical Assessment} (\emph{NLOS IMT-A UMi}
model - \emph{Scenario 2}, $T=32$, $R=16$, $SNR=20$ {[}dB{]}, $P=100$).
Behaviour of (\emph{a}) $\mathcal{C}_{\chi,r}$, (\emph{b}) $\mathcal{I}_{\chi,r}$,
and (\emph{c}) $\mathcal{D}_{\chi,r}$ versus the $r$-th ($r=1,...,R$)
receiver index ($\chi\in\left\{ V,H\right\} $).
\item \textbf{Figure 8.} \emph{Numerical Assessment} (\emph{NLOS IMT-A UMi}
model - \emph{Scenario 2}, $T=32$, $R=16$, $SNR=20$ {[}dB{]}, $r=19$,
$p=1$). Plots of (\emph{a})(\emph{b}) the magnitude and (\emph{c})(\emph{d})
the phase of $w_{\chi,\psi,r,t}^{\left(p\right)}$ versus the $t$-th
($t=1,...,T$) element index, and of (\emph{e})(\emph{f}) the normalized
magnitude of $A_{\chi,r}^{\left(p\right),\mathcal{C}}\left(\theta,\varphi\right)$,
$A_{\chi,r}^{\left(p\right),\mathcal{I}}\left(\theta,\varphi\right)$,
and $A_{\chi,r}^{\left(p\right)}\left(\theta,\varphi\right)$ along
the $\theta=90$ {[}deg{]}-cut when (\emph{a})(\emph{c})(\emph{e})
$\chi=\psi=V$ or (\emph{b})(\emph{d})(\emph{f}) $\chi=\psi=H$.
\item \textbf{Figure 9.} \emph{Numerical Assessment} (\emph{NLOS IMT-A UMi}
model - \emph{Scenario 2}, $T=32$, $R=16$, $SNR=20$ {[}dB{]}, $r=19$,
$p=1$, $\chi=\psi=V$). Normalized 3D magnitude of (\emph{a}) $A_{\chi,r}^{\left(p\right),\mathcal{C}}\left(\theta,\varphi\right)$,
(\emph{b}) $A_{\chi,r}^{\left(p\right),\mathcal{I}}\left(\theta,\varphi\right)$,
and (\emph{c}) $A_{\chi,r}^{\left(p\right)}\left(\theta,\varphi\right)$.
\item \textbf{Figure 10.} \emph{Numerical Assessment} (\emph{NLOS IMT-A
UMi} model - \emph{Scenario 1}, $R=16$, $SNR=20$ {[}dB{]}). Behaviour
of (\emph{a}) $\mathcal{C}_{ave}$, (\emph{b}) $\mathcal{I}_{ave}$,
and (\emph{c}) $\mathcal{D}_{ave}$ \emph{}versus $T$. 
\item \textbf{Figure 11.} \emph{Numerical Assessment} (\emph{NLOS IMT-A
UMi} model - \emph{Scenario 1}, $T=72$, $SNR=20$ {[}dB{]}). Behaviour
of (\emph{a}) $\mathcal{C}_{ave}$, (\emph{b}) $\mathcal{I}_{ave}$,
and (\emph{c}) $\mathcal{D}_{ave}$ versus $R$.
\item \textbf{Figure 12.} \emph{Numerical Assessment} ($S=32$, $L=16$).
Behaviour of (\emph{a}) $\mathcal{C}_{ave}$ and (\emph{b}) $\mathcal{I}_{ave}$
versus $SNR$.
\item \textbf{Figure 13.} \emph{Numerical Assessment} (\emph{NLOS IMT-A
UMi} model - \emph{Scenario 1}, $SNR=20$ {[}dB{]}). Behaviour of
$\tau^{ZF}$, $\tau^{ISO}$, and $\tau^{HCS}$ (\emph{a}) versus $T$
($R=16$) and (\emph{b}) versus $R$ ($T=72$).
\end{itemize}

\section*{TABLE CAPTIONS}

\begin{itemize}
\item \textbf{Table I.} \emph{Numerical Assessment} (\emph{NLOS IMT-A UMi}
model - \emph{Scenario 1}, $T=32$, $R=16$, $SNR=20$ {[}dB{]}, $P=100$).
Average performance indexes.
\item \textbf{Table II.} \emph{Numerical Assessment} (\emph{NLOS IMT-A UMi}
model - \emph{Scenario 2}, $T=32$, $R=16$, $SNR=20$ {[}dB{]}, $P=100$).
Average performance indexes.\newpage

\end{itemize}
\begin{center}~\vfill\end{center}

\begin{center}\includegraphics[%
  width=0.90\columnwidth]{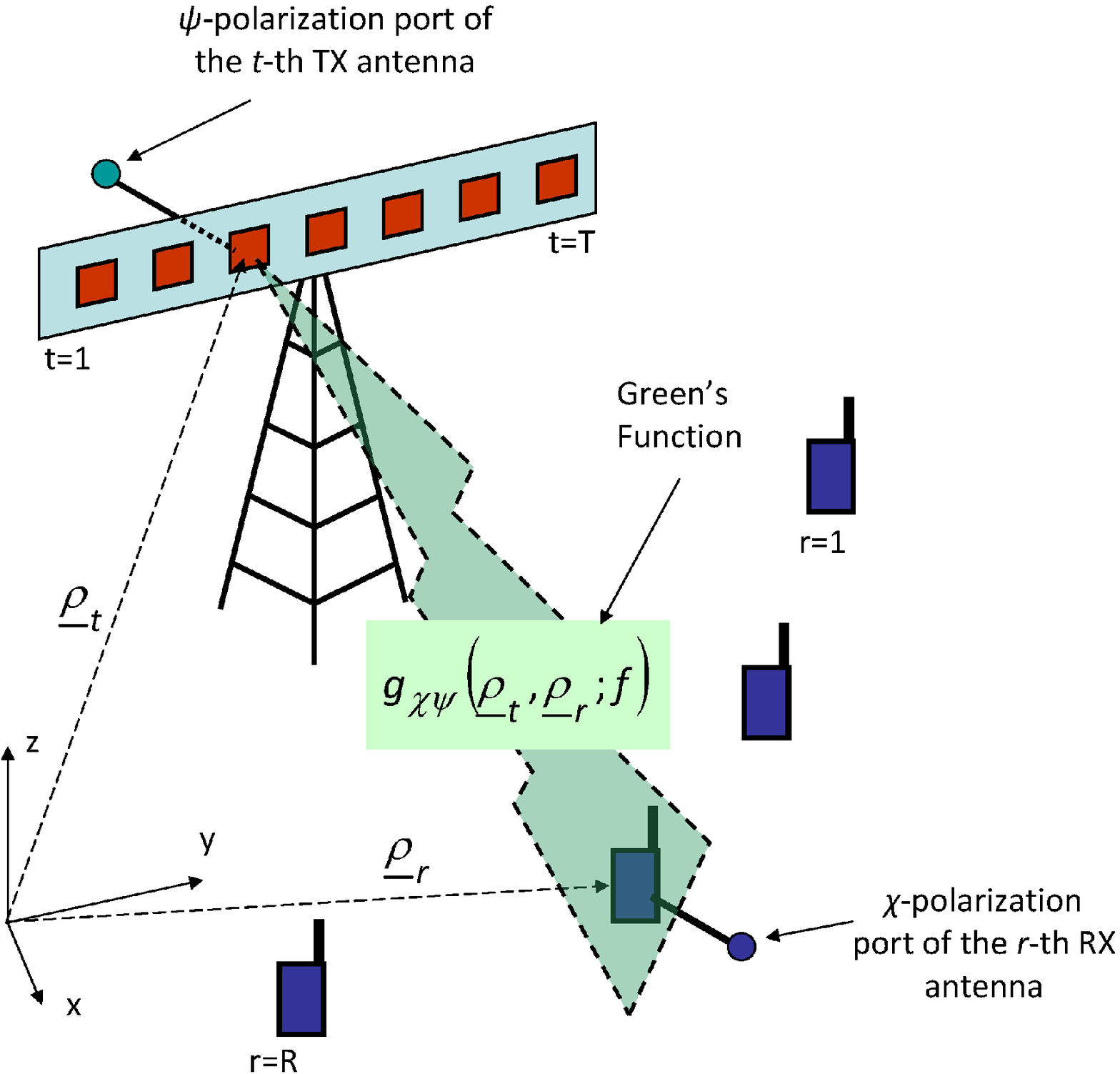}\end{center}

\begin{center}\vfill\end{center}

\begin{center}\textbf{Figure 1 - G. Oliveri et} \textbf{\emph{al.}},
{}``Capacity-Driven Low-Interference Fast Beam Synthesis ...''\end{center}
\newpage

\begin{center}\vfill\end{center}

\begin{center}\begin{tabular}{cc}
\multicolumn{2}{c}{\includegraphics[%
  width=0.55\columnwidth,
  keepaspectratio]{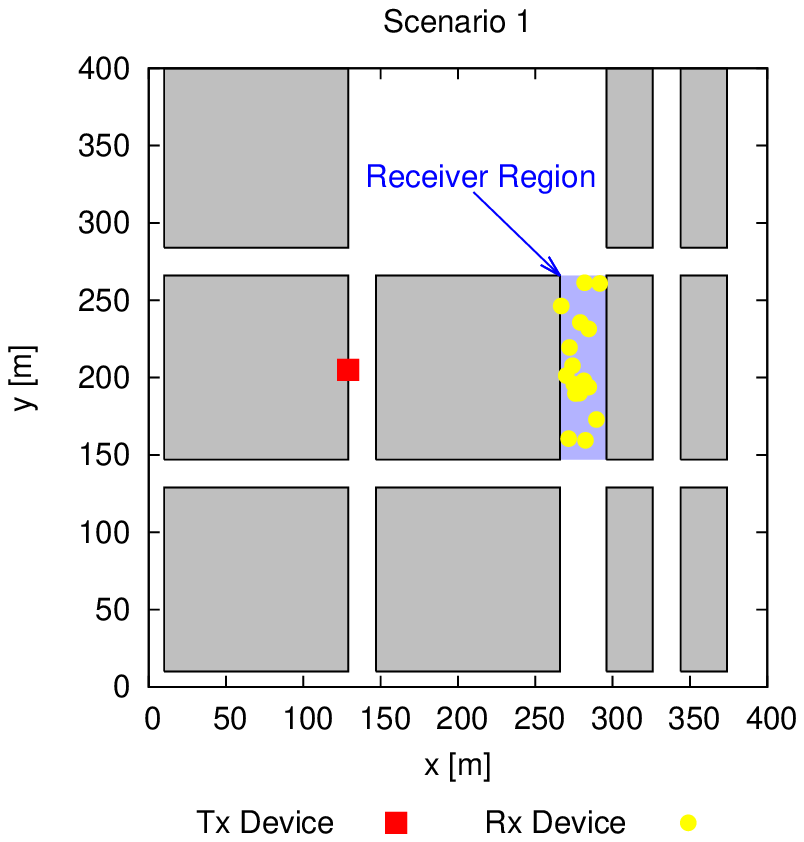}}\tabularnewline
\multicolumn{2}{c}{(\emph{a})}\tabularnewline
\multicolumn{2}{c}{\includegraphics[%
  width=0.55\columnwidth,
  keepaspectratio]{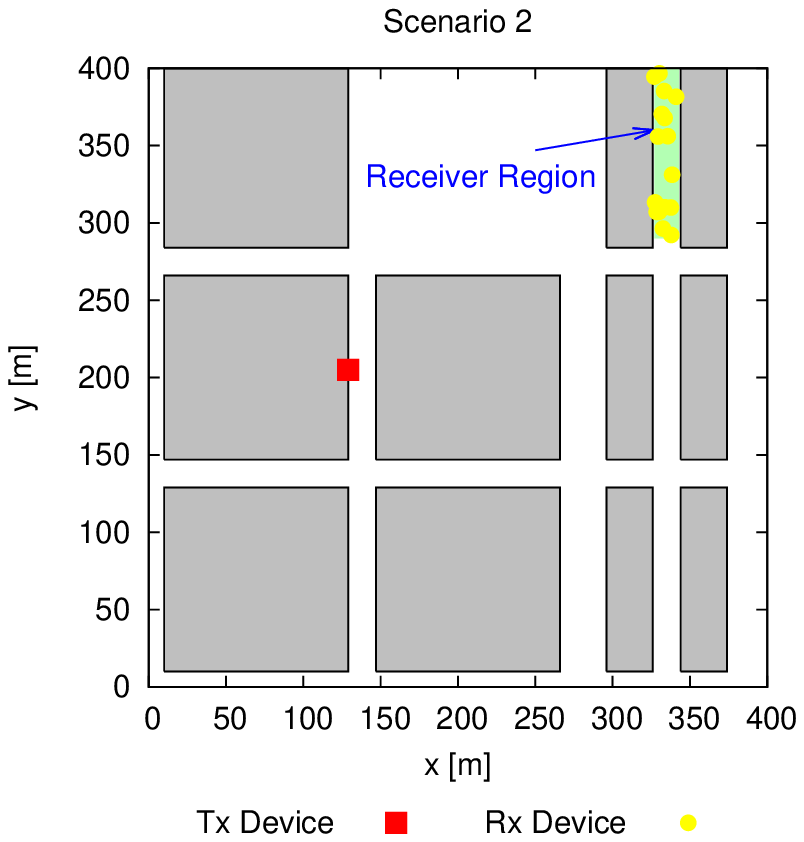}}\tabularnewline
\multicolumn{2}{c}{(\emph{b})}\tabularnewline
\end{tabular}\end{center}

\begin{center}\vfill\end{center}

\begin{center}\textbf{Figure 2 - G. Oliveri et} \textbf{\emph{al.}},
{}``Capacity-Driven Low-Interference Fast Beam Synthesis ...''\end{center}
\newpage

\begin{center}\begin{tabular}{cc}
\multicolumn{2}{c}{\includegraphics[%
  width=0.60\columnwidth,
  keepaspectratio]{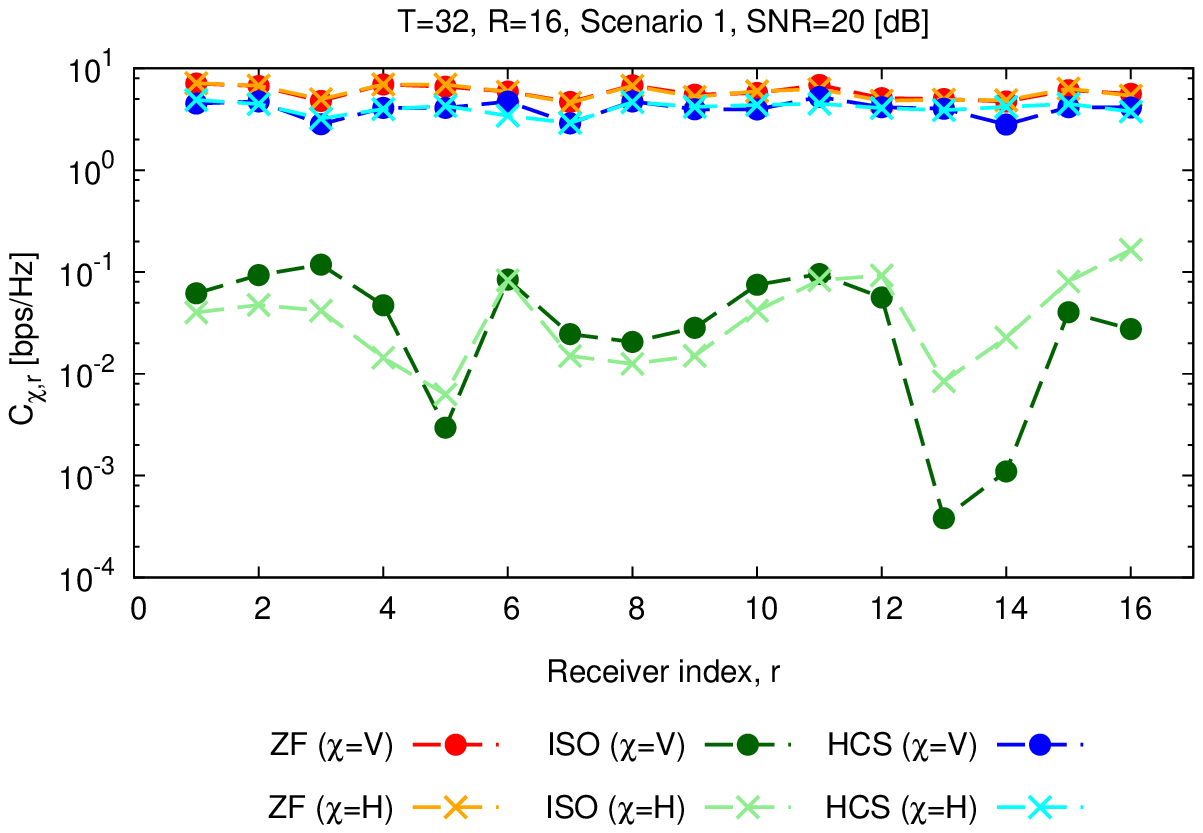}}\tabularnewline
\multicolumn{2}{c}{(\emph{a})}\tabularnewline
\multicolumn{2}{c}{\includegraphics[%
  width=0.60\columnwidth,
  keepaspectratio]{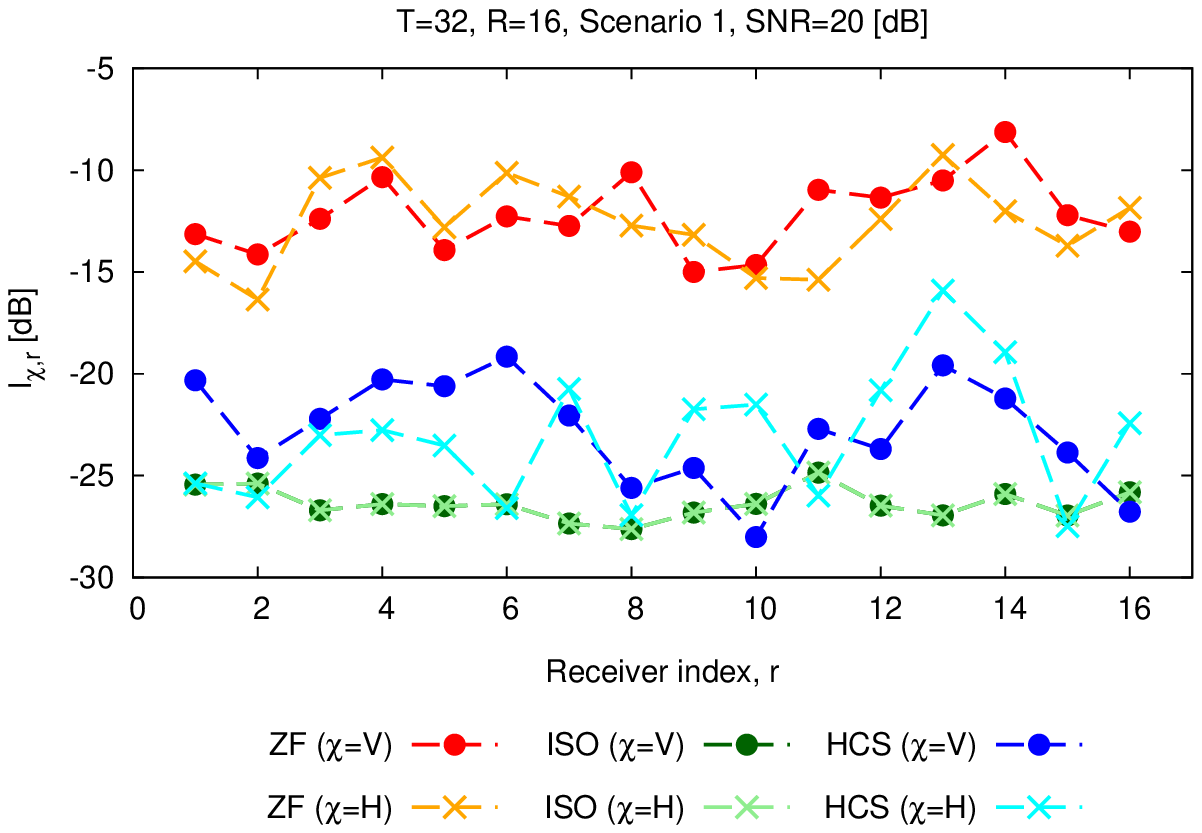}}\tabularnewline
\multicolumn{2}{c}{(\emph{b})}\tabularnewline
\multicolumn{2}{c}{\includegraphics[%
  width=0.60\columnwidth,
  keepaspectratio]{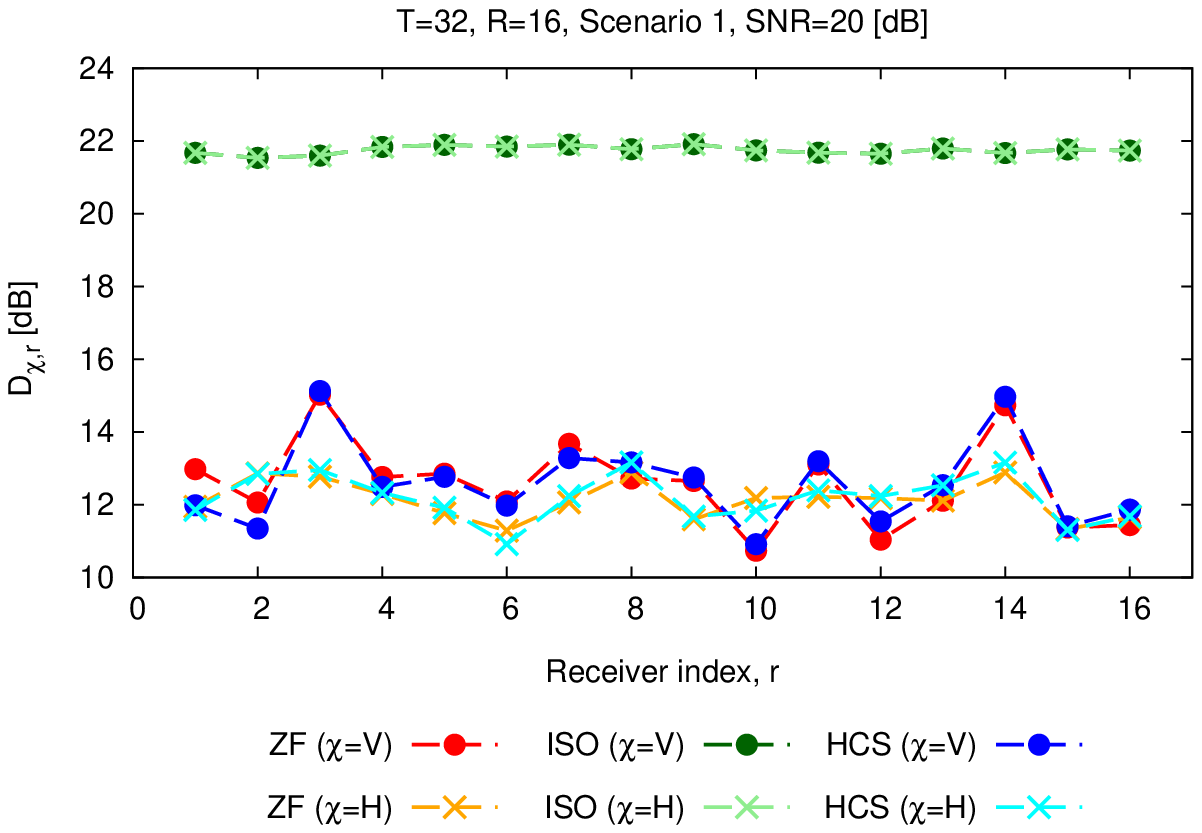}}\tabularnewline
\multicolumn{2}{c}{(\emph{c})}\tabularnewline
\end{tabular}\end{center}

\begin{center}\textbf{Figure 3 - G. Oliveri et} \textbf{\emph{al.}},
{}``Capacity-Driven Low-Interference Fast Beam Synthesis ...''\end{center}
\newpage

\begin{center}~\vfill\end{center}

\begin{center}\begin{tabular}{cc}
\includegraphics[%
  width=0.45\columnwidth,
  keepaspectratio]{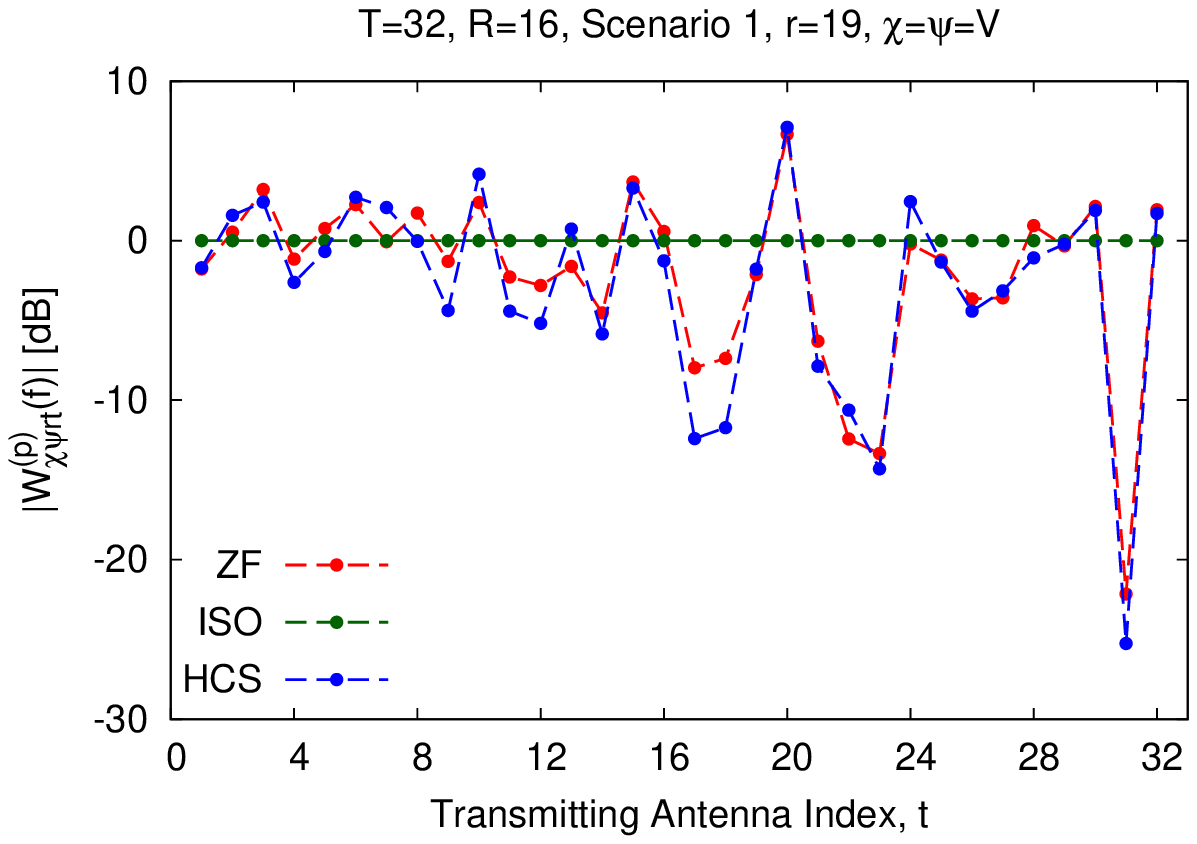}&
\includegraphics[%
  width=0.45\columnwidth,
  keepaspectratio]{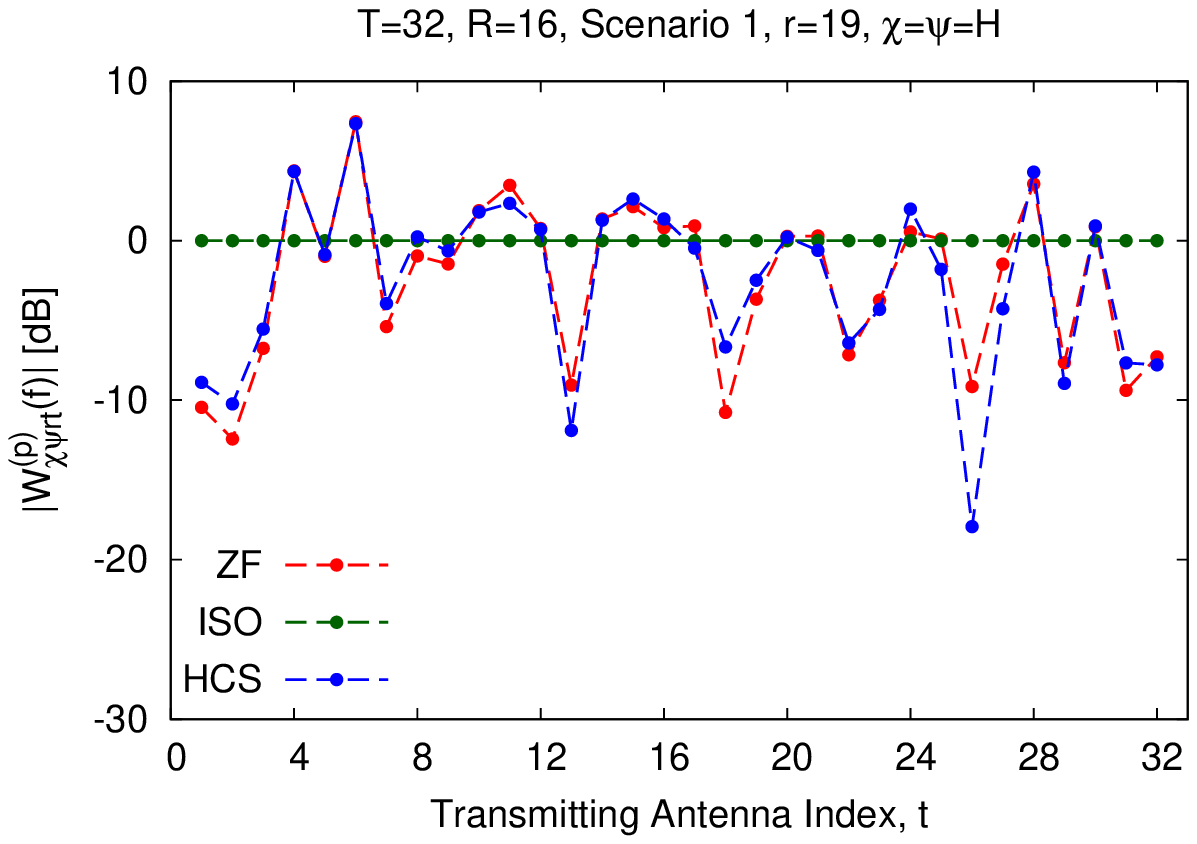}\tabularnewline
(\emph{a})&
(\emph{b})\tabularnewline
\includegraphics[%
  width=0.45\columnwidth,
  keepaspectratio]{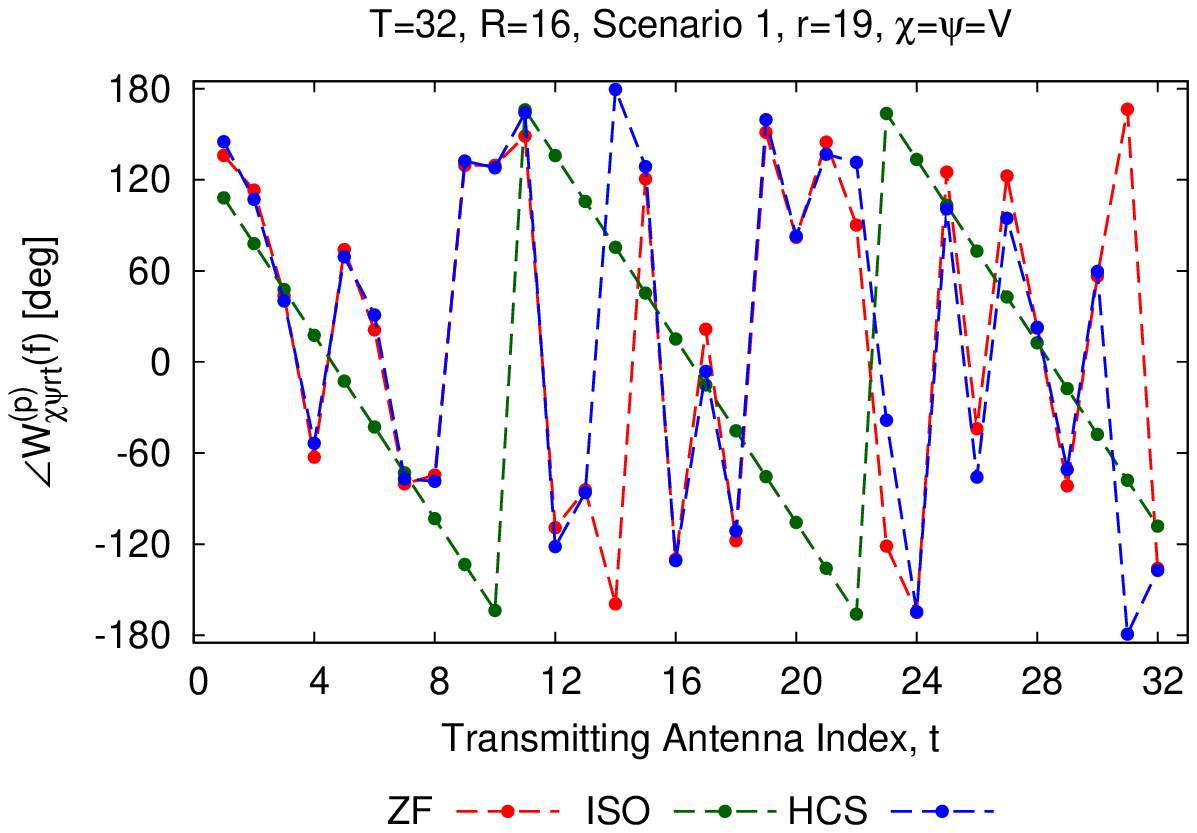}&
\includegraphics[%
  width=0.45\columnwidth,
  keepaspectratio]{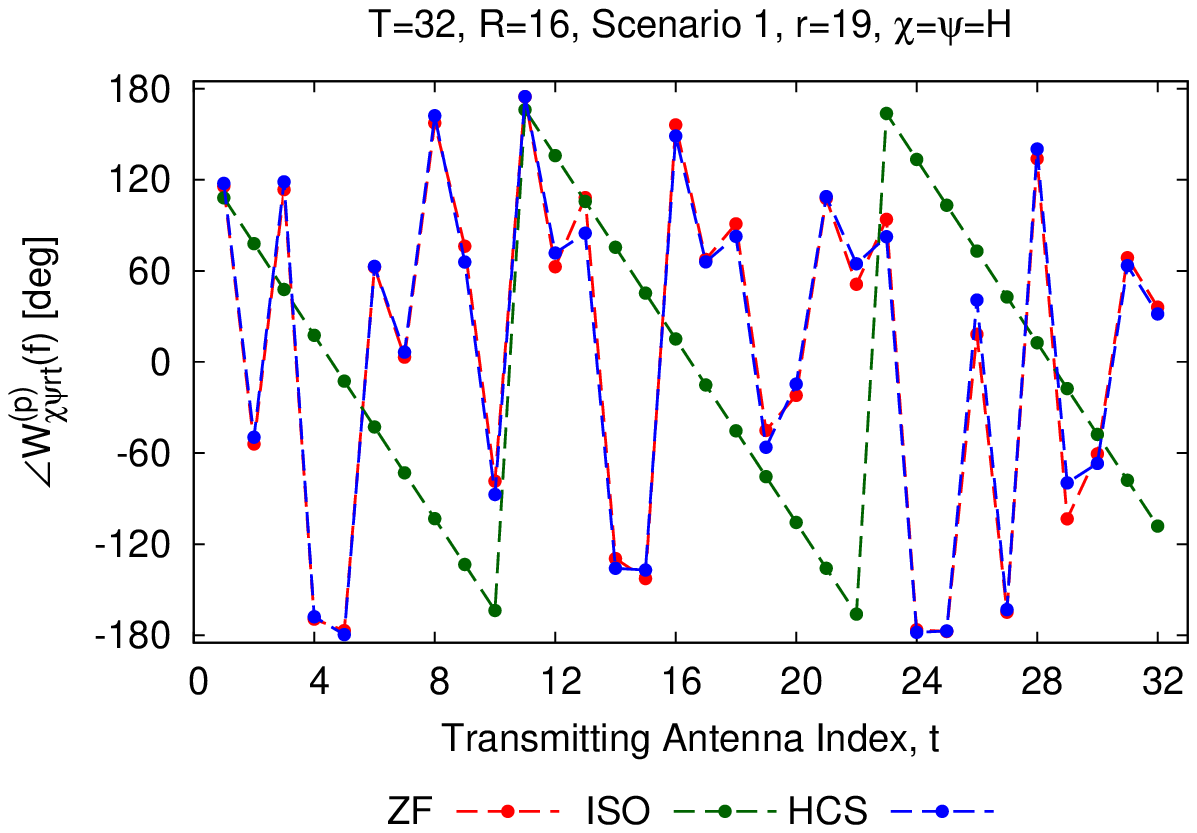}\tabularnewline
(\emph{c})&
(\emph{d})\tabularnewline
\end{tabular}\end{center}

\begin{center}\vfill\end{center}

\begin{center}\textbf{Figure 4 - G. Oliveri et} \textbf{\emph{al.}},
{}``Capacity-Driven Low-Interference Fast Beam Synthesis ...''\end{center}
\newpage

\begin{center}~\vfill\end{center}

\begin{center}\begin{tabular}{cc}
\multicolumn{2}{c}{\includegraphics[%
  width=0.85\columnwidth,
  keepaspectratio]{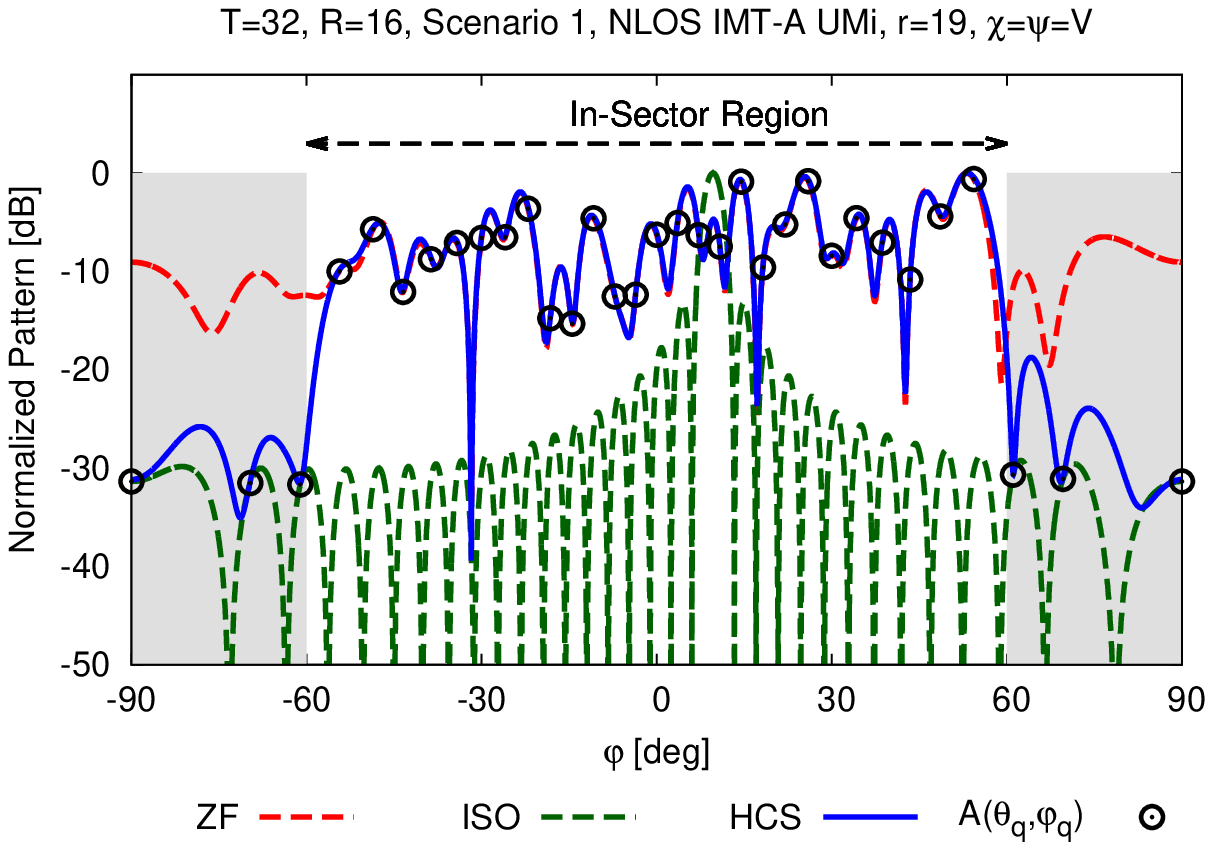}}\tabularnewline
\multicolumn{2}{c}{(\emph{a})}\tabularnewline
\multicolumn{2}{c}{\includegraphics[%
  width=0.85\columnwidth,
  keepaspectratio]{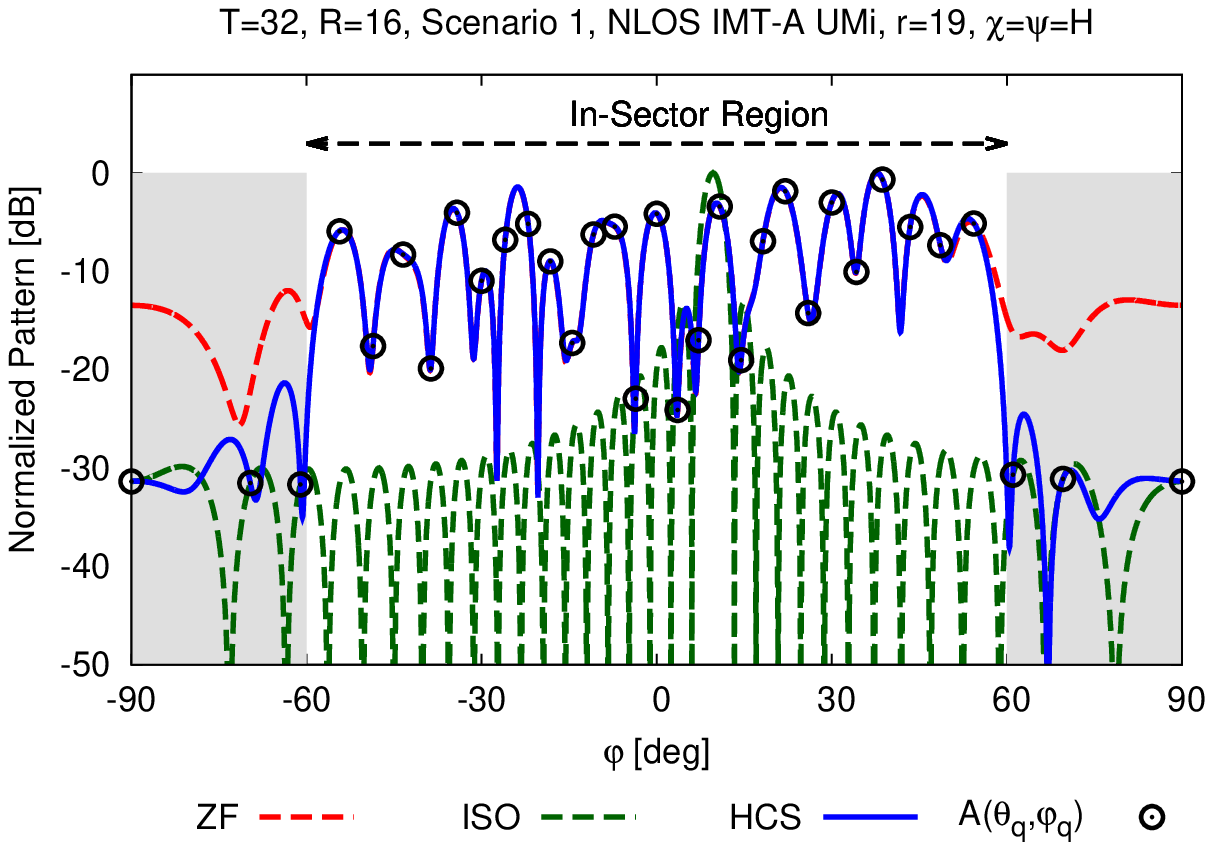}}\tabularnewline
\multicolumn{2}{c}{(\emph{b})}\tabularnewline
\end{tabular}\end{center}

\begin{center}\vfill\end{center}

\begin{center}\textbf{Figure 5 - G. Oliveri et} \textbf{\emph{al.}},
{}``Capacity-Driven Low-Interference Fast Beam Synthesis ...''\end{center}
\newpage

\begin{center}\begin{tabular}{c}
\includegraphics[%
  width=0.45\columnwidth,
  keepaspectratio]{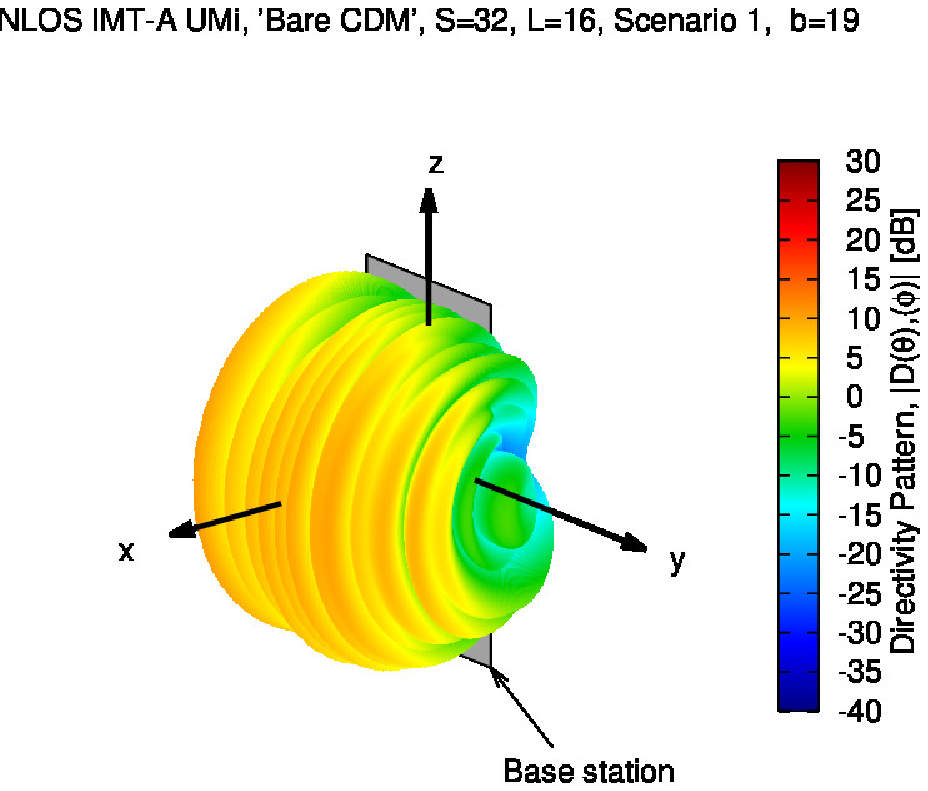}\tabularnewline
(\emph{a})\tabularnewline
\includegraphics[%
  width=0.45\columnwidth,
  keepaspectratio]{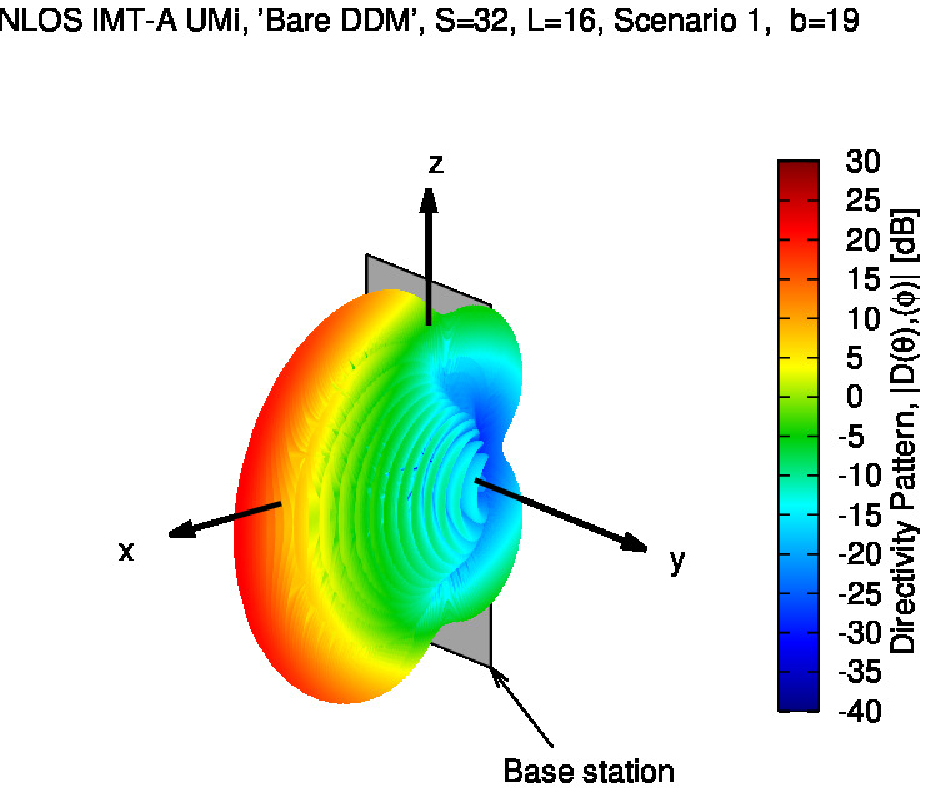}\tabularnewline
(\emph{b})\tabularnewline
\includegraphics[%
  width=0.45\columnwidth,
  keepaspectratio]{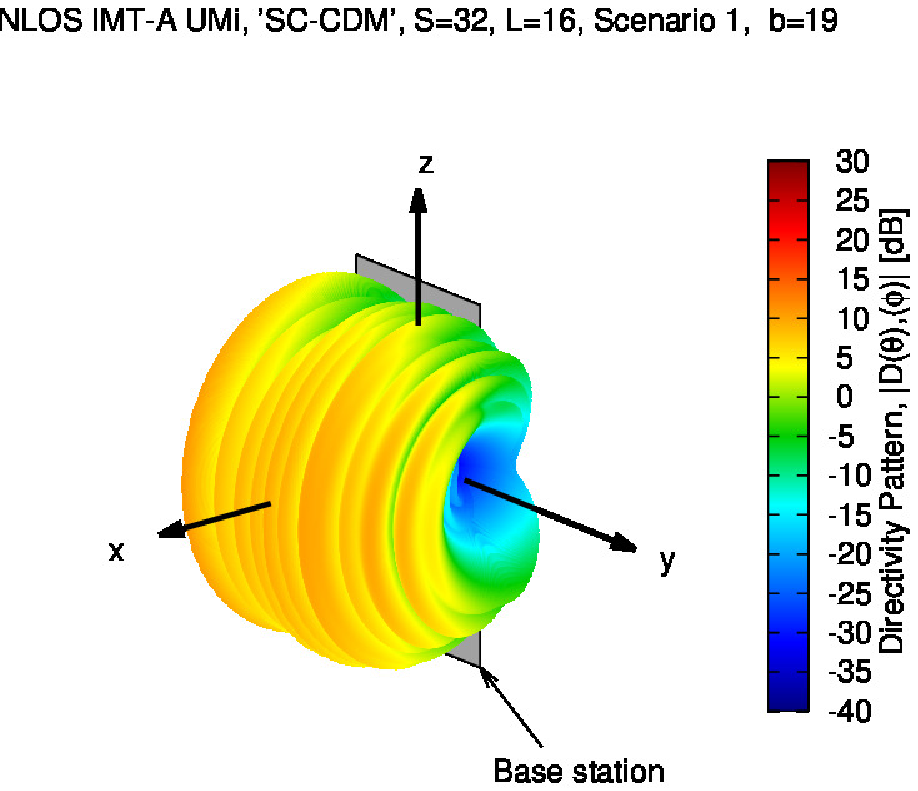}\tabularnewline
(\emph{c})\tabularnewline
\end{tabular}\end{center}

\begin{center}\vfill\end{center}

\begin{center}\textbf{Figure 6 - G. Oliveri et} \textbf{\emph{al.}},
{}``Capacity-Driven Low-Interference Fast Beam Synthesis ...''\end{center}
\newpage

\begin{center}\begin{tabular}{cc}
\multicolumn{2}{c}{\includegraphics[%
  width=0.60\columnwidth,
  keepaspectratio]{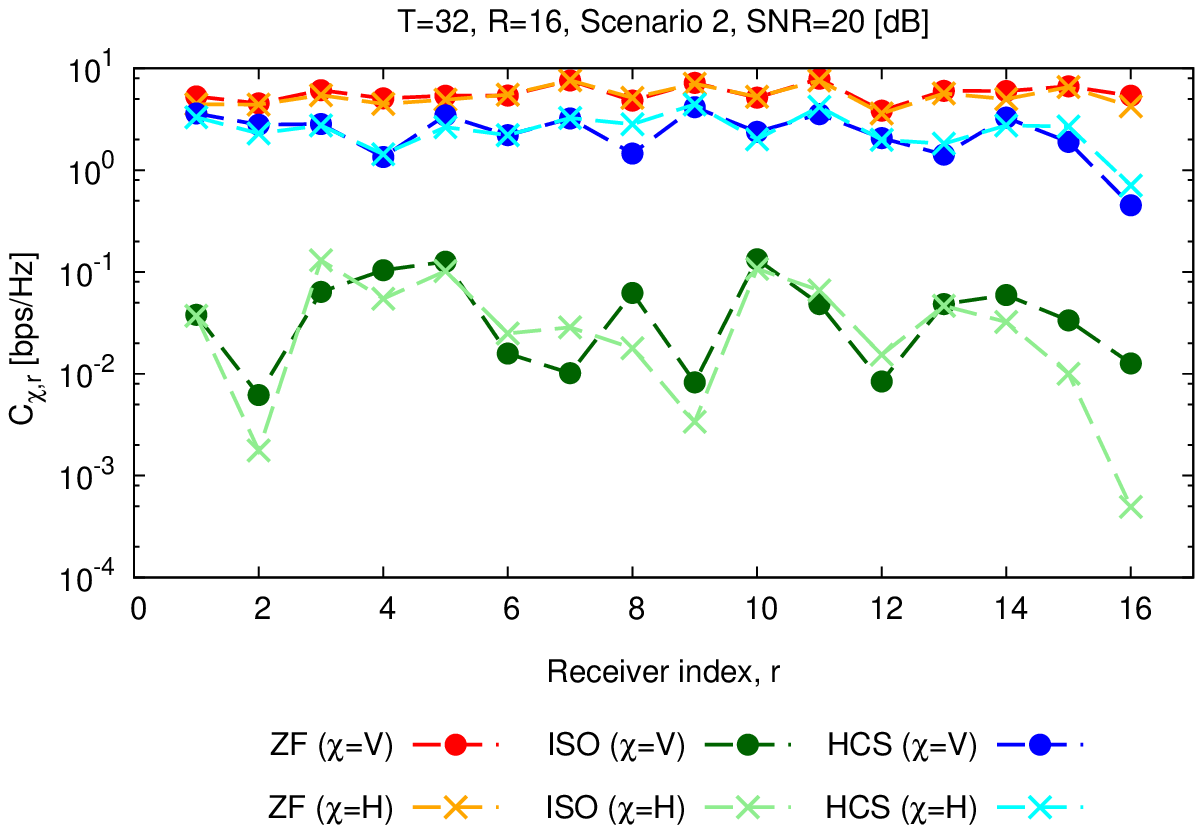}}\tabularnewline
\multicolumn{2}{c}{(\emph{a})}\tabularnewline
\multicolumn{2}{c}{\includegraphics[%
  width=0.60\columnwidth,
  keepaspectratio]{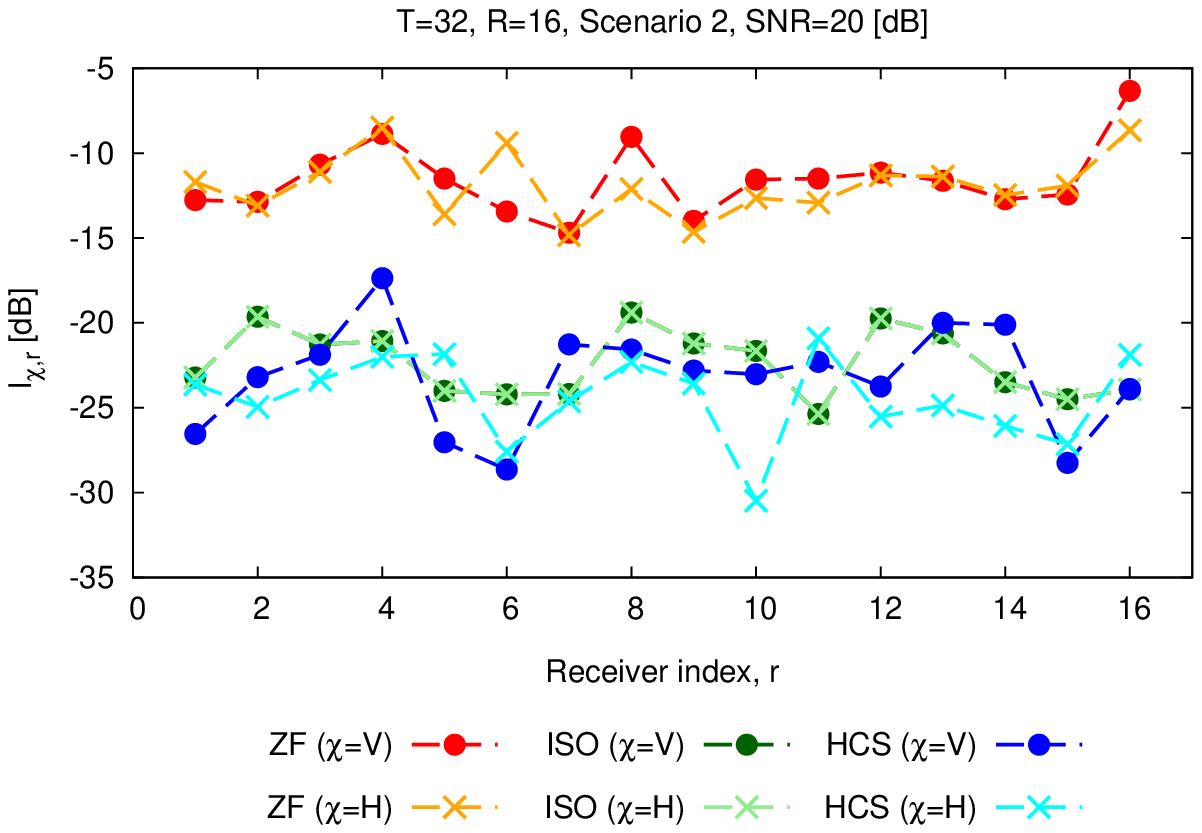}}\tabularnewline
\multicolumn{2}{c}{(\emph{b})}\tabularnewline
\multicolumn{2}{c}{\includegraphics[%
  width=0.60\columnwidth,
  keepaspectratio]{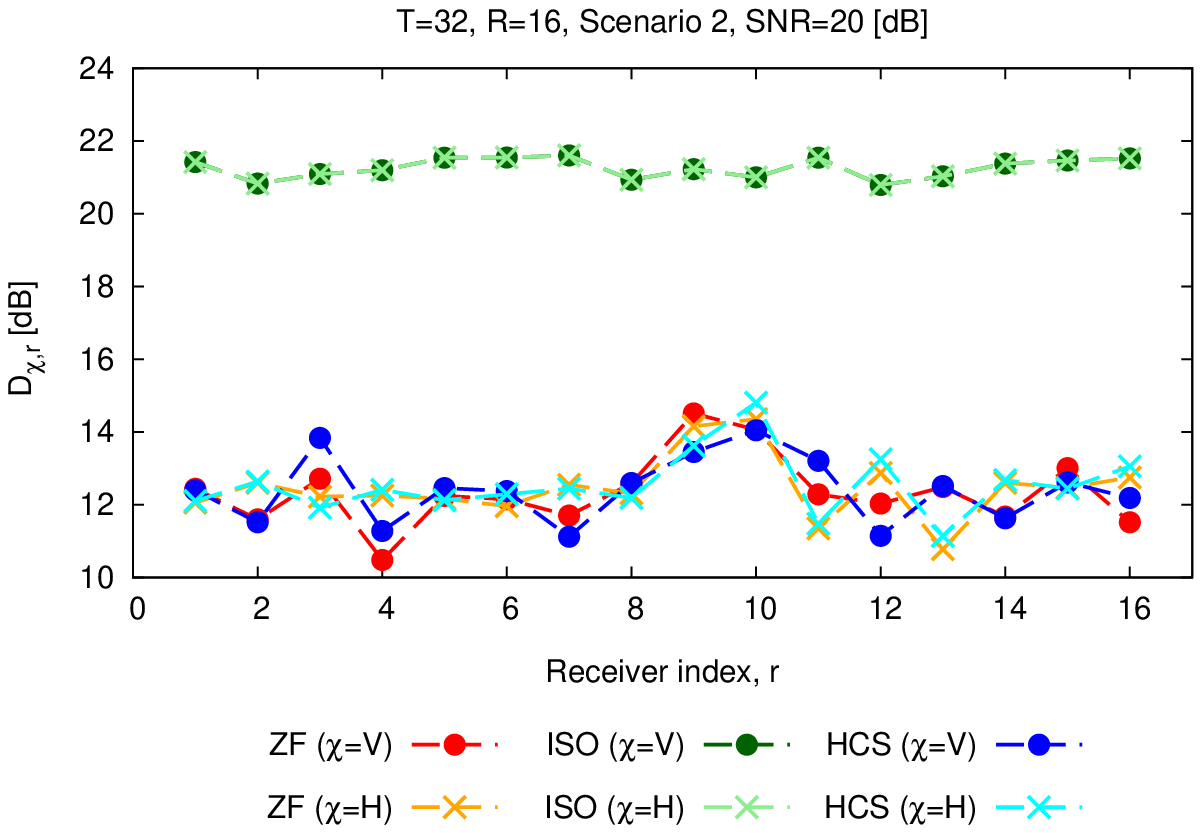}}\tabularnewline
\multicolumn{2}{c}{(\emph{c})}\tabularnewline
\end{tabular}\end{center}

\begin{center}\textbf{Figure 7 - G. Oliveri et} \textbf{\emph{al.}},
{}``Capacity-Driven Low-Interference Fast Beam Synthesis ...''\end{center}
\newpage

\begin{center}~\vfill\end{center}

\begin{center}\begin{tabular}{cc}
\includegraphics[%
  width=0.45\columnwidth,
  keepaspectratio]{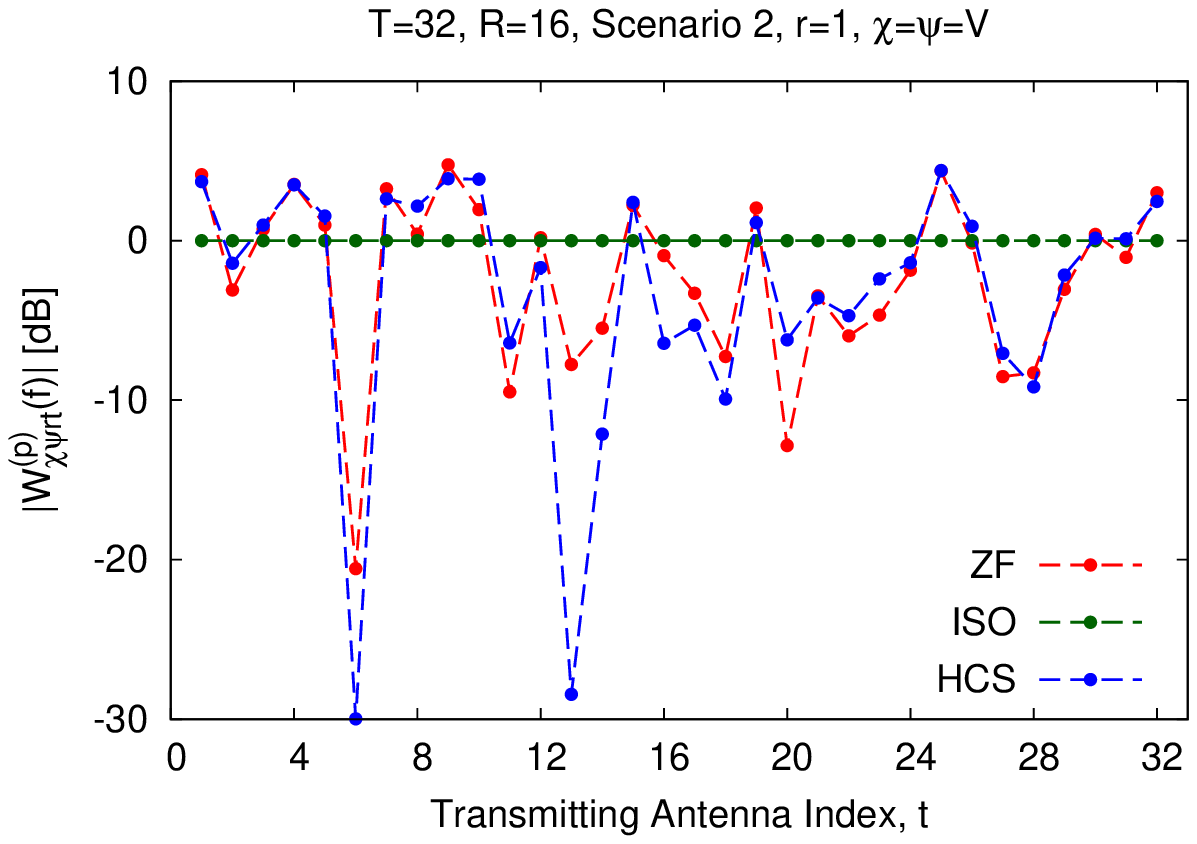}&
\includegraphics[%
  width=0.45\columnwidth,
  keepaspectratio]{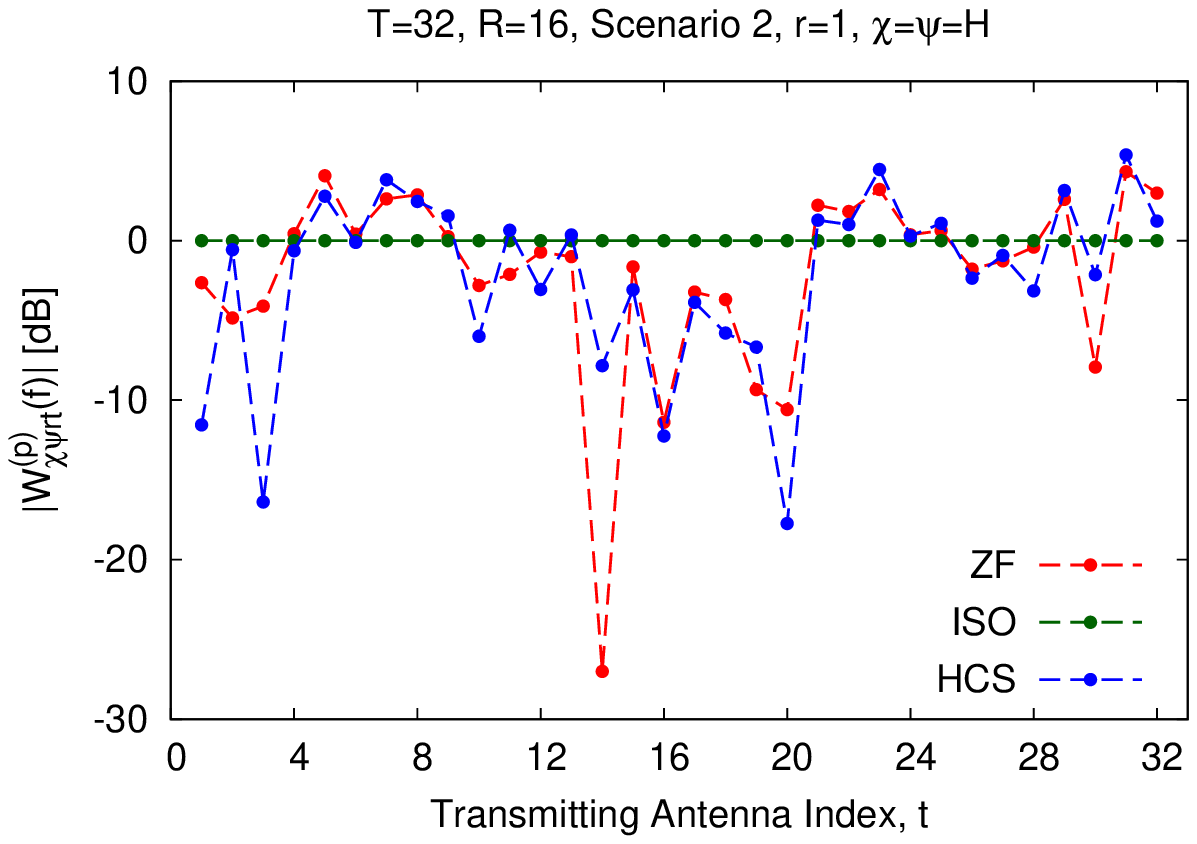}\tabularnewline
(\emph{a})&
(\emph{b})\tabularnewline
\includegraphics[%
  width=0.45\columnwidth,
  keepaspectratio]{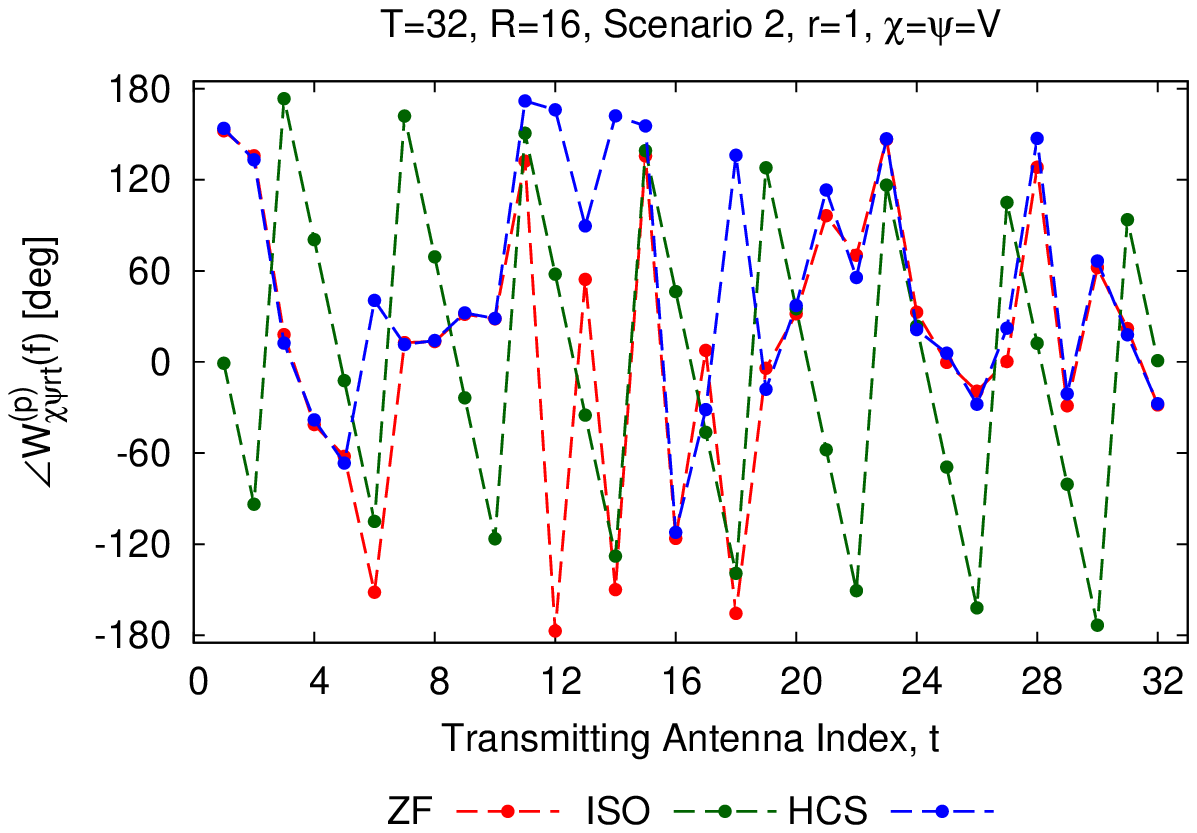}&
\includegraphics[%
  width=0.45\columnwidth,
  keepaspectratio]{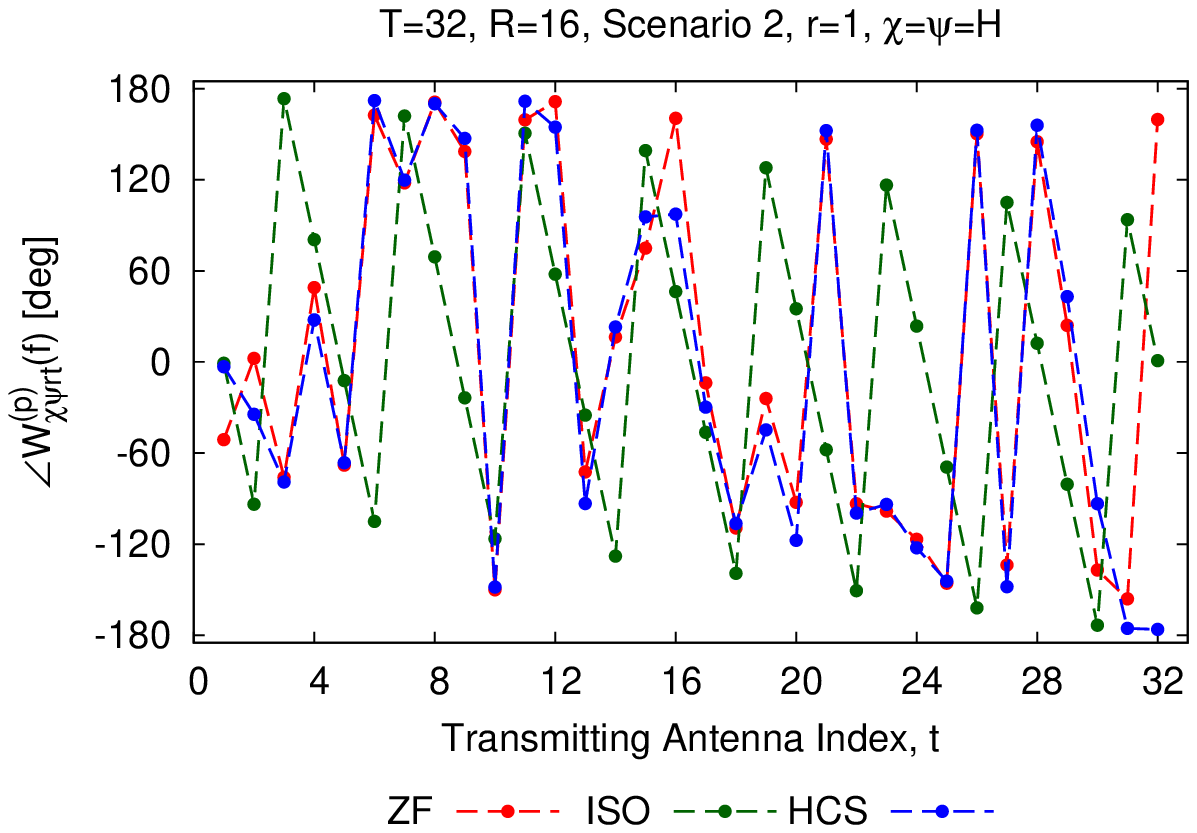}\tabularnewline
(\emph{c})&
(\emph{d})\tabularnewline
\includegraphics[%
  width=0.45\columnwidth,
  keepaspectratio]{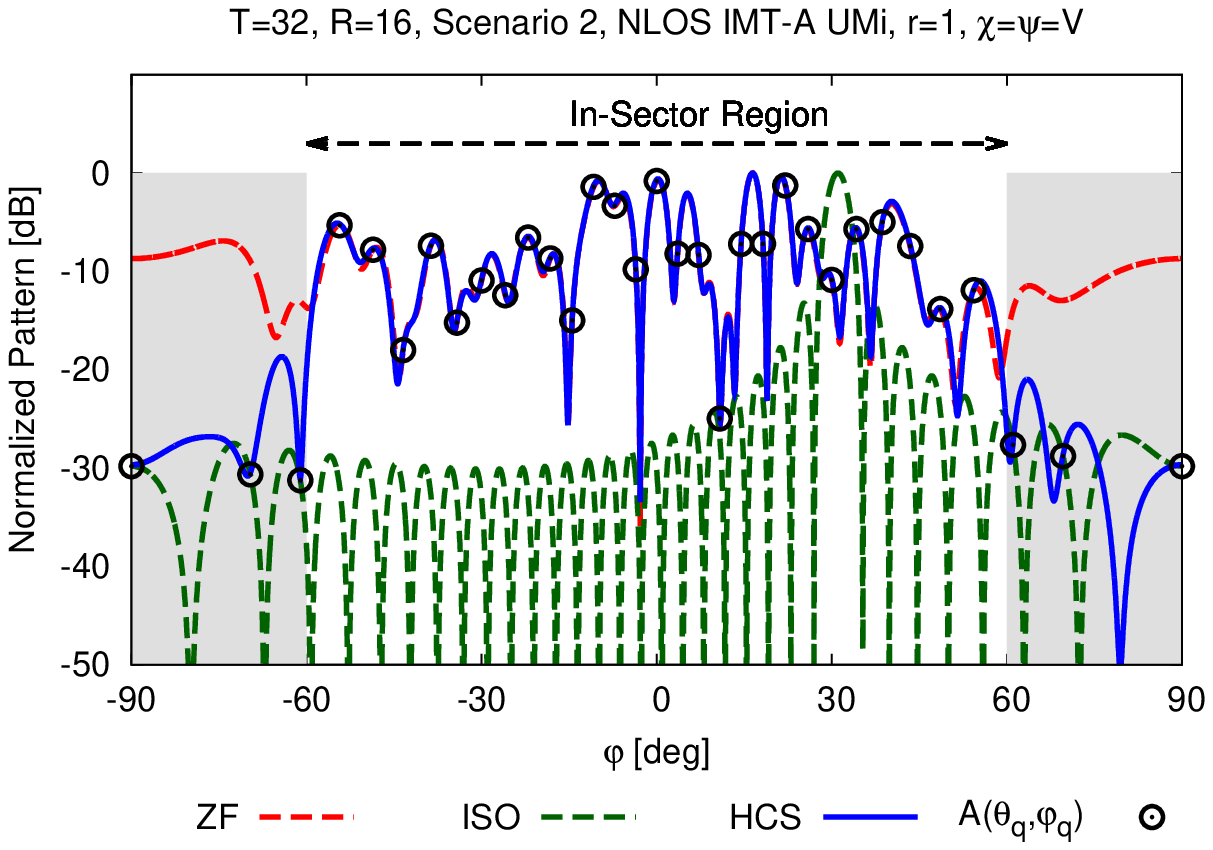}&
\includegraphics[%
  width=0.45\columnwidth,
  keepaspectratio]{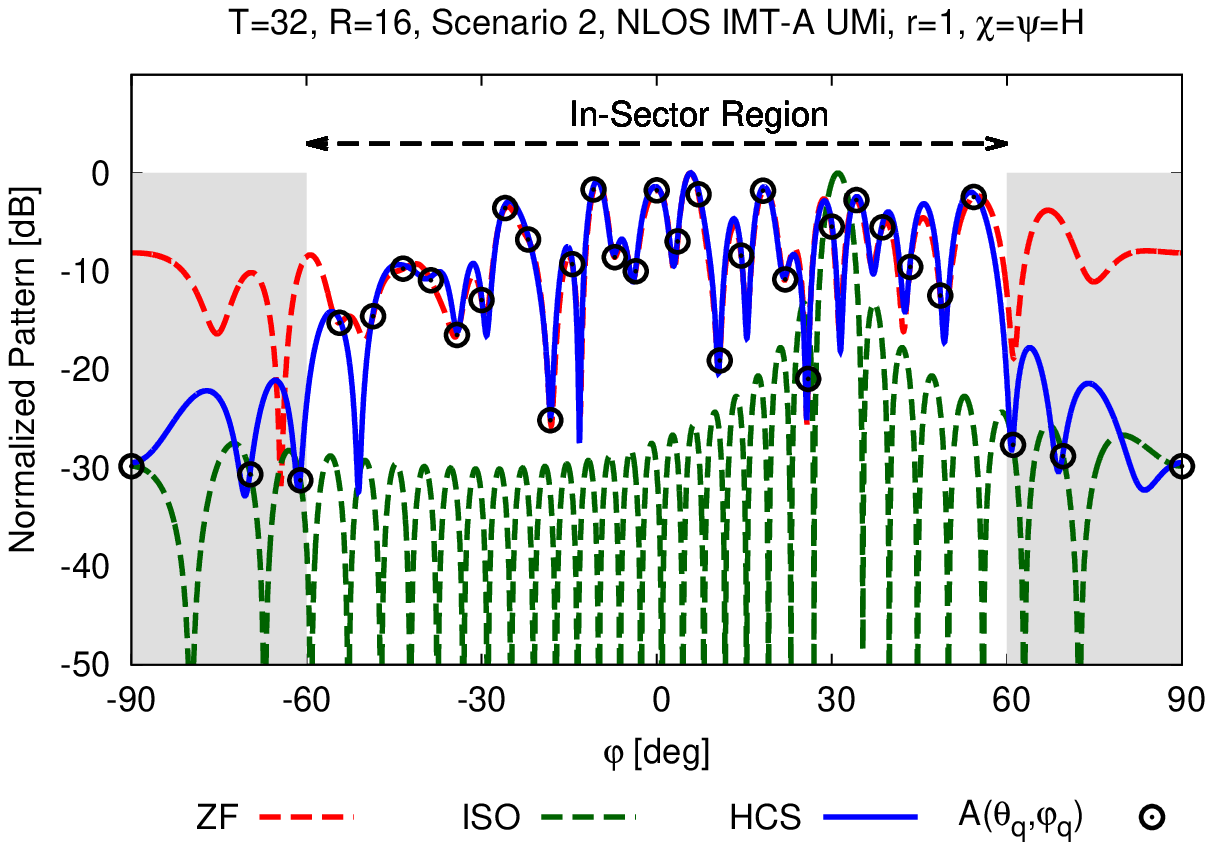}\tabularnewline
(\emph{e})&
(\emph{f})\tabularnewline
\end{tabular}\end{center}

\begin{center}\vfill\end{center}

\begin{center}\textbf{Figure 8 - G. Oliveri et} \textbf{\emph{al.}},
{}``Capacity-Driven Low-Interference Fast Beam Synthesis ...''\end{center}
\newpage

\begin{center}\begin{tabular}{c}
\includegraphics[%
  width=0.45\columnwidth,
  keepaspectratio]{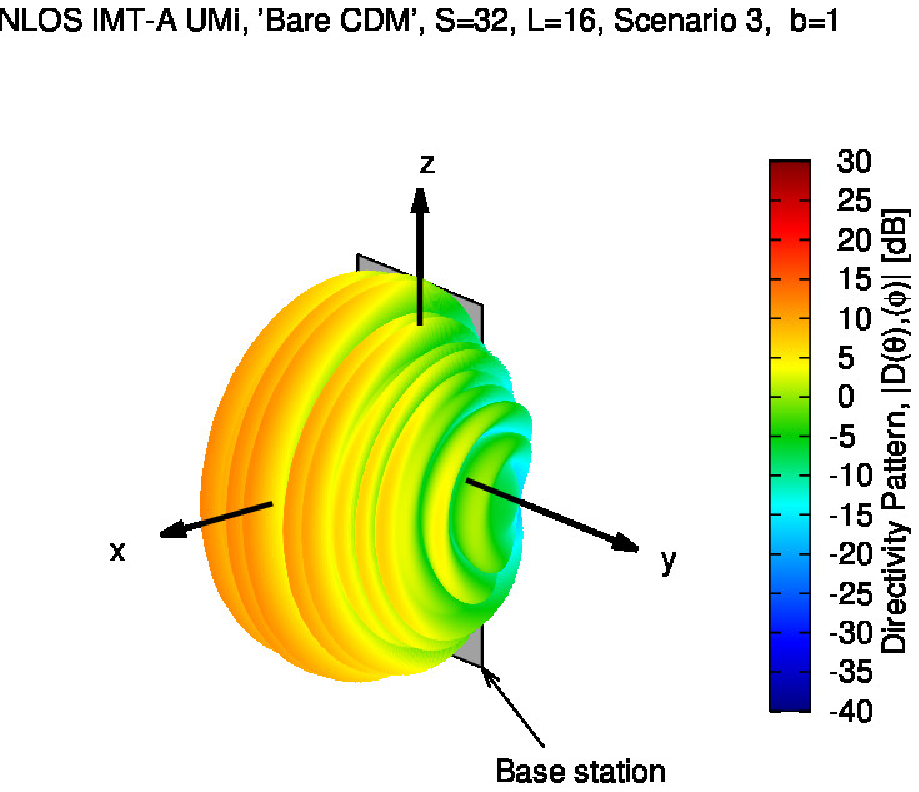}\tabularnewline
(\emph{a})\tabularnewline
\includegraphics[%
  width=0.45\columnwidth,
  keepaspectratio]{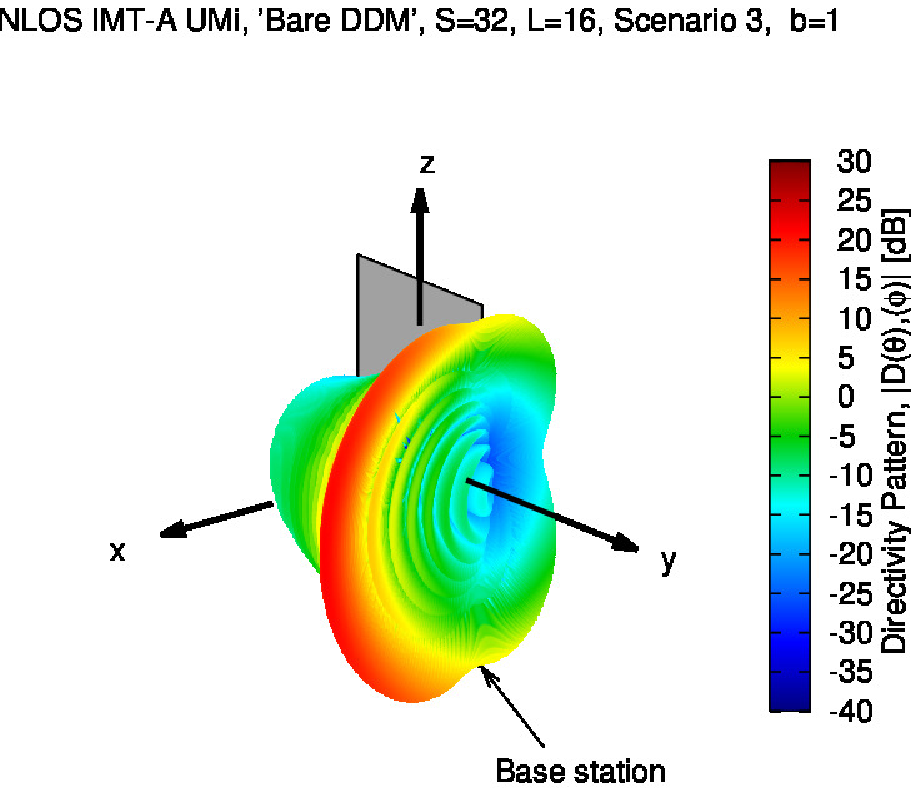}\tabularnewline
(\emph{b})\tabularnewline
\includegraphics[%
  width=0.45\columnwidth,
  keepaspectratio]{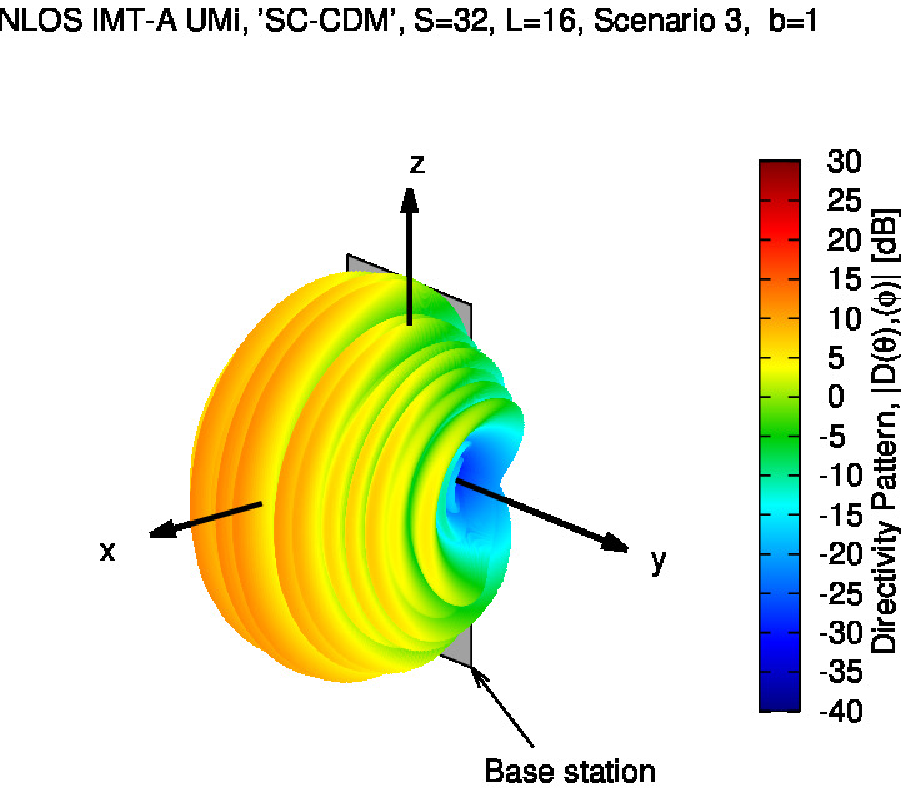}\tabularnewline
(\emph{c})\tabularnewline
\end{tabular}\end{center}

\begin{center}\vfill\end{center}

\begin{center}\textbf{Figure 9 - G. Oliveri et} \textbf{\emph{al.}},
{}``Capacity-Driven Low-Interference Fast Beam Synthesis ...''\end{center}
\newpage

\begin{center}\begin{tabular}{cc}
\multicolumn{2}{c}{\includegraphics[%
  width=0.60\columnwidth,
  keepaspectratio]{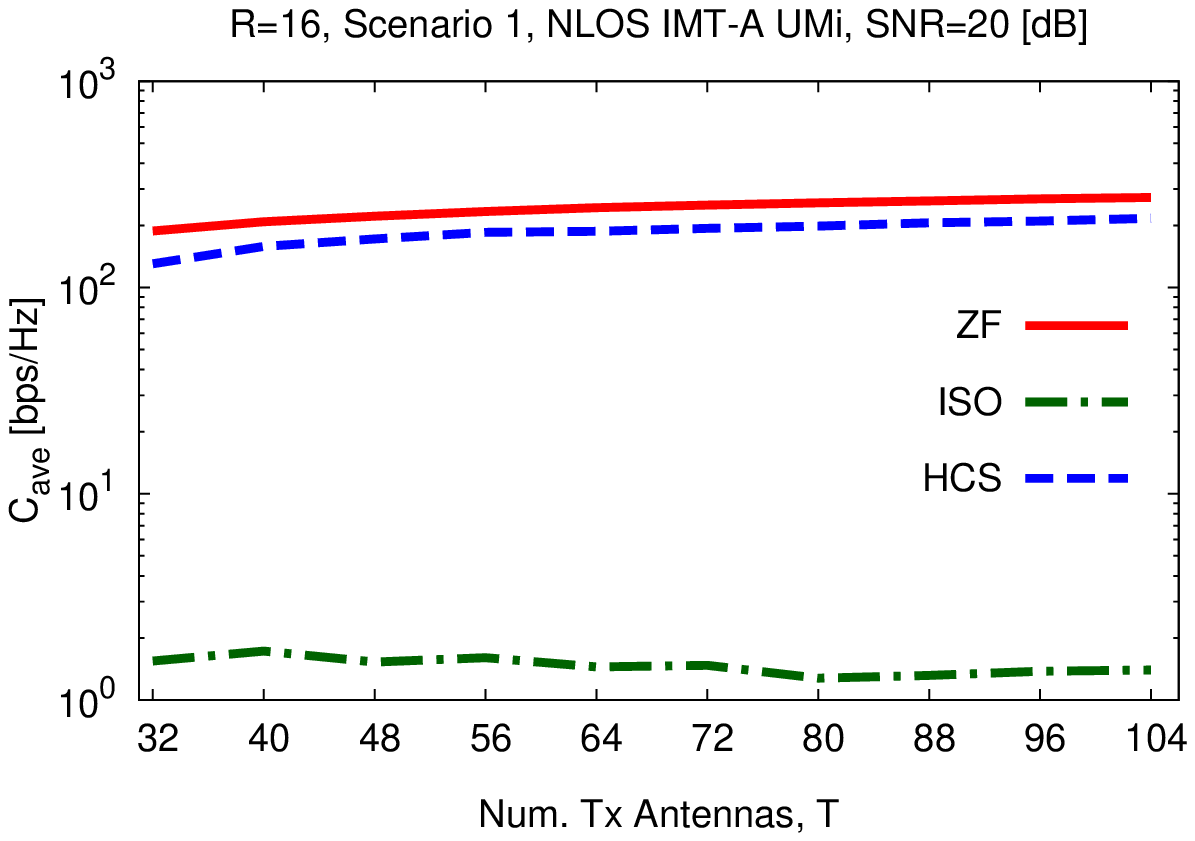}}\tabularnewline
\multicolumn{2}{c}{(\emph{a})}\tabularnewline
\multicolumn{2}{c}{\includegraphics[%
  width=0.60\columnwidth,
  keepaspectratio]{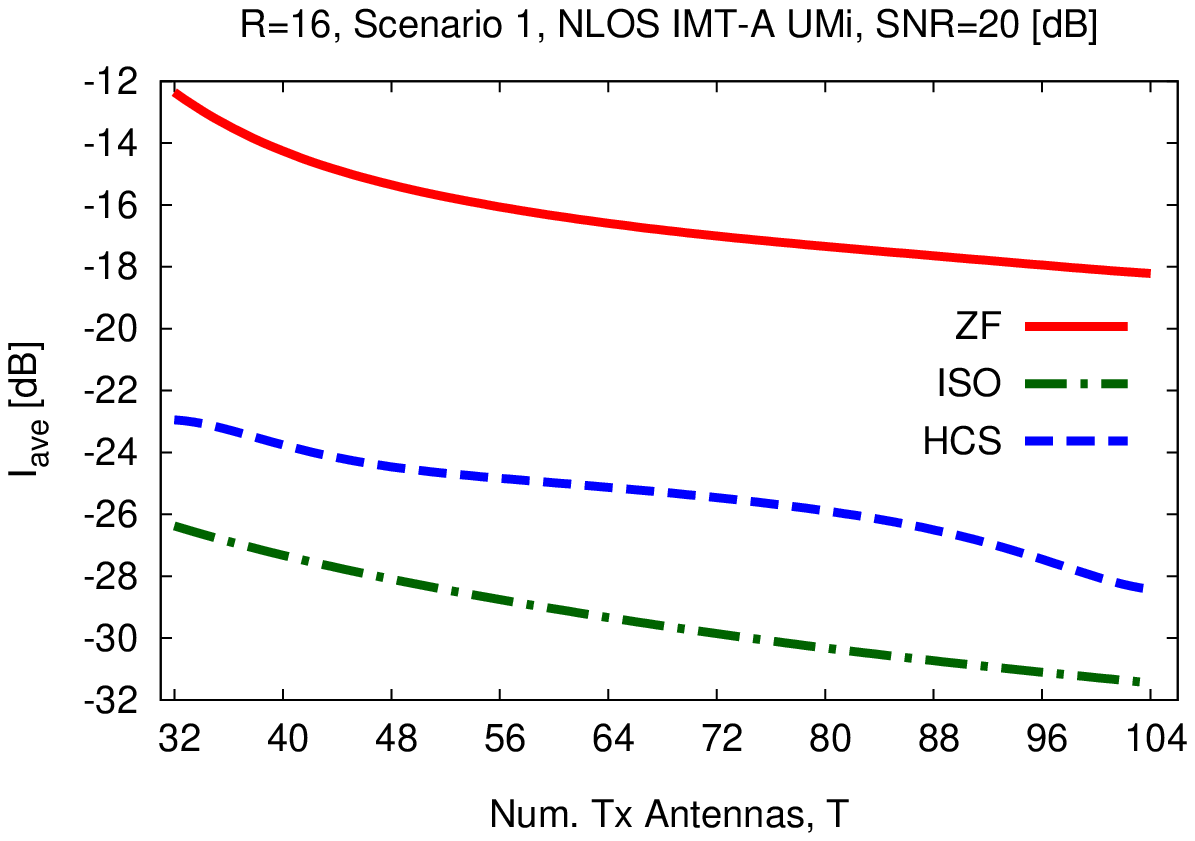}}\tabularnewline
\multicolumn{2}{c}{(\emph{b})}\tabularnewline
\multicolumn{2}{c}{\includegraphics[%
  width=0.60\columnwidth,
  keepaspectratio]{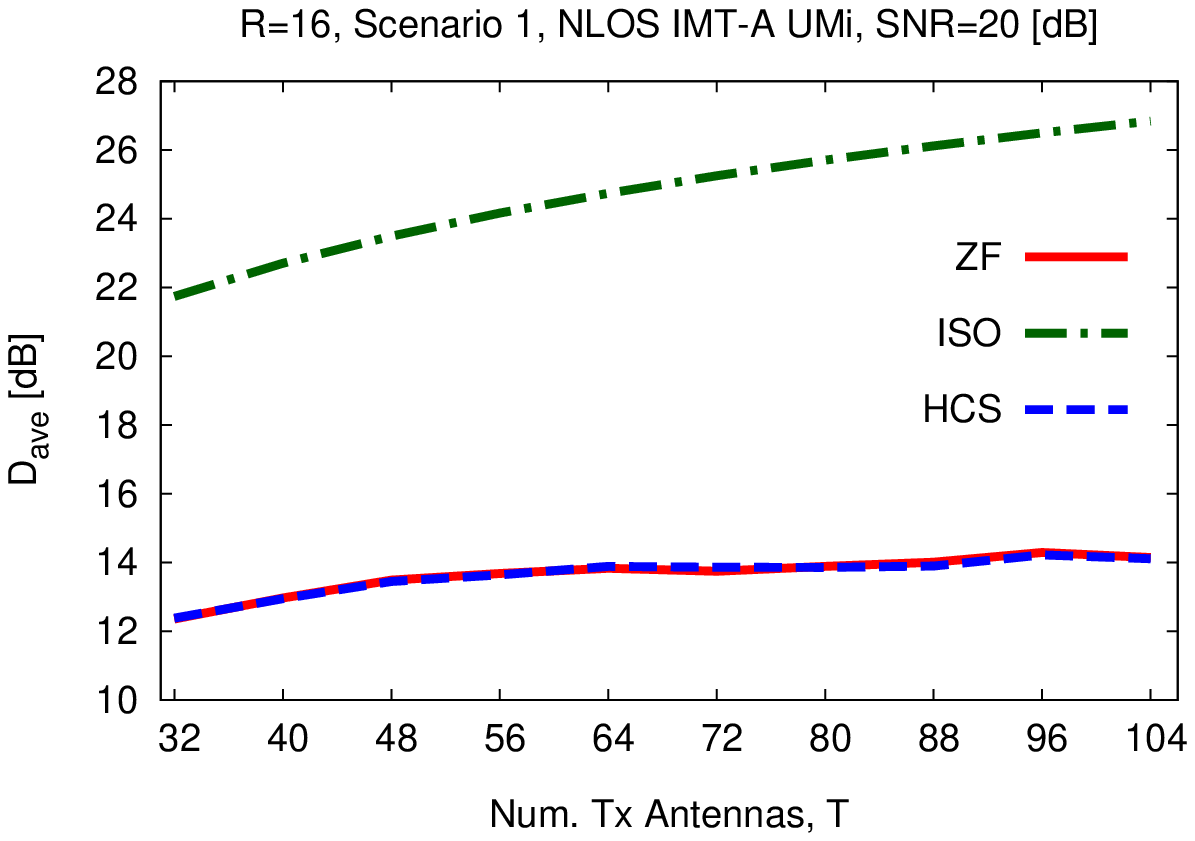}}\tabularnewline
\multicolumn{2}{c}{(\emph{c})}\tabularnewline
\end{tabular}\end{center}

\begin{center}\textbf{Figure 10 - G. Oliveri et} \textbf{\emph{al.}},
{}``Capacity-Driven Low-Interference Fast Beam Synthesis ...''\end{center}
\newpage

\begin{center}\begin{tabular}{cc}
\multicolumn{2}{c}{\includegraphics[%
  width=0.60\columnwidth,
  keepaspectratio]{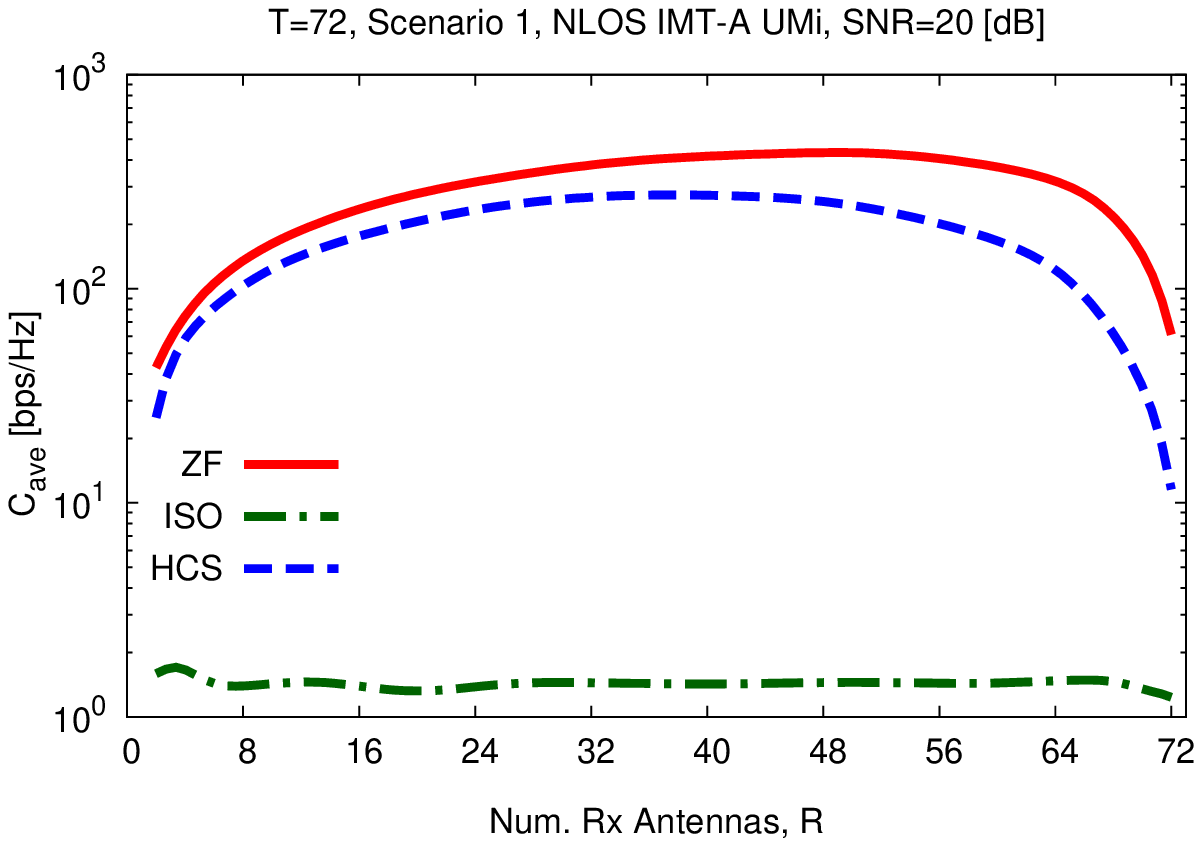}}\tabularnewline
\multicolumn{2}{c}{(\emph{a})}\tabularnewline
\multicolumn{2}{c}{\includegraphics[%
  width=0.60\columnwidth,
  keepaspectratio]{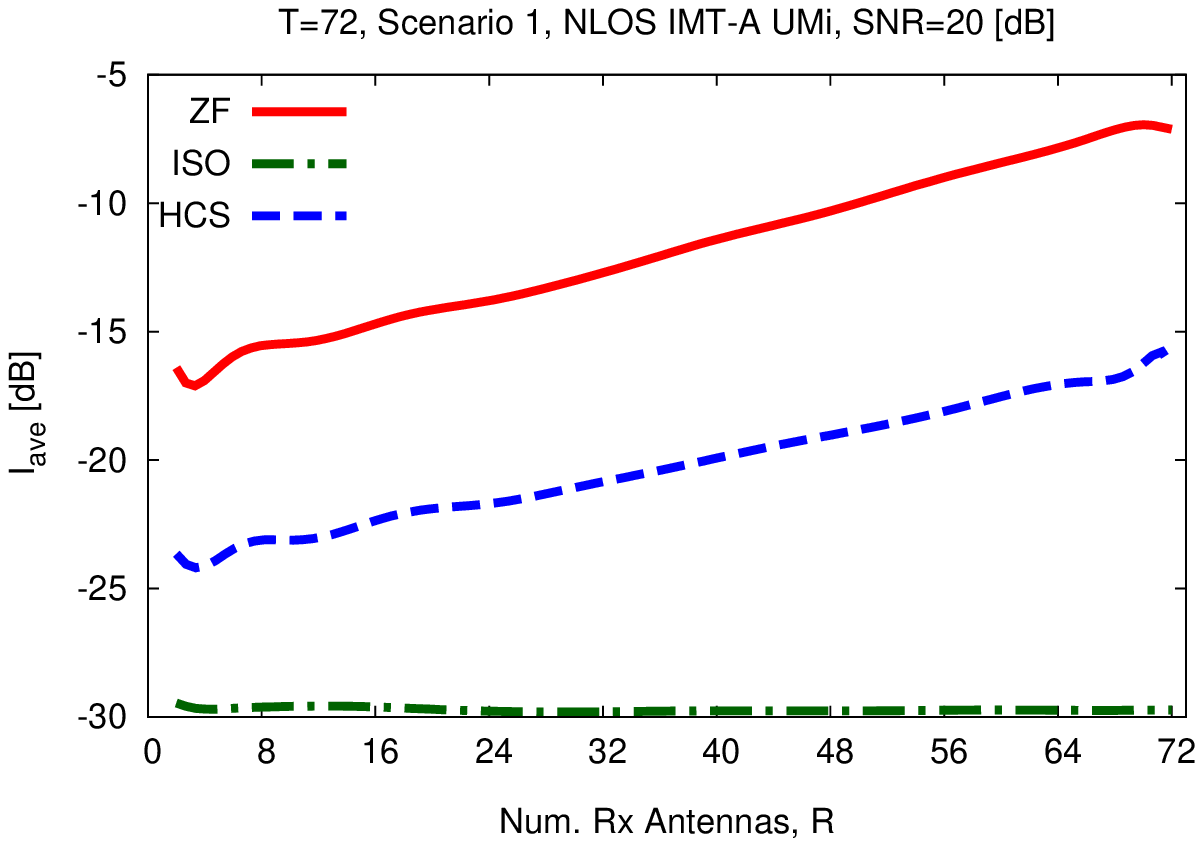}}\tabularnewline
\multicolumn{2}{c}{(\emph{b})}\tabularnewline
\multicolumn{2}{c}{\includegraphics[%
  width=0.60\columnwidth,
  keepaspectratio]{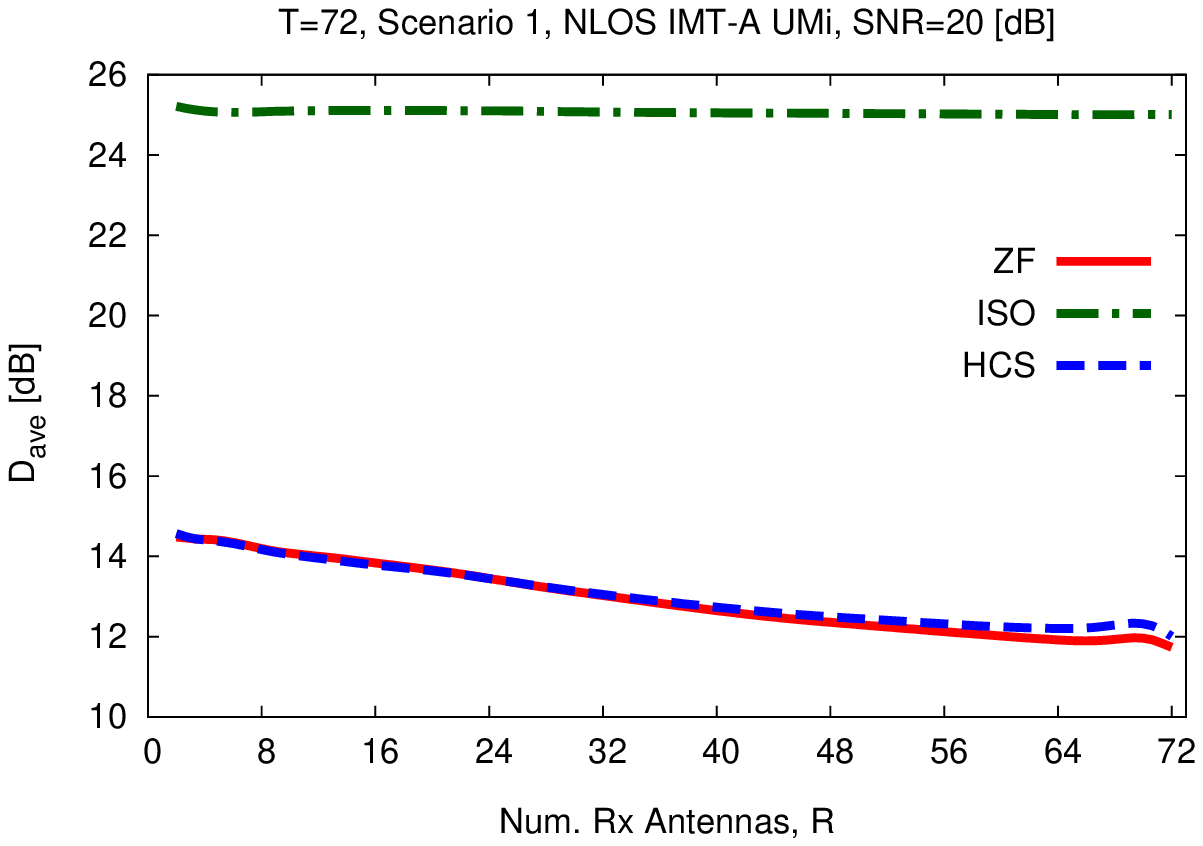}}\tabularnewline
\multicolumn{2}{c}{(\emph{c})}\tabularnewline
\end{tabular}\end{center}

\begin{center}\textbf{Figure 11 - G. Oliveri et} \textbf{\emph{al.}},
{}``Capacity-Driven Low-Interference Fast Beam Synthesis ...''\end{center}
\newpage

\begin{center}~\vfill\end{center}

\begin{center}\begin{tabular}{cc}
\multicolumn{2}{c}{\includegraphics[%
  width=0.95\columnwidth,
  keepaspectratio]{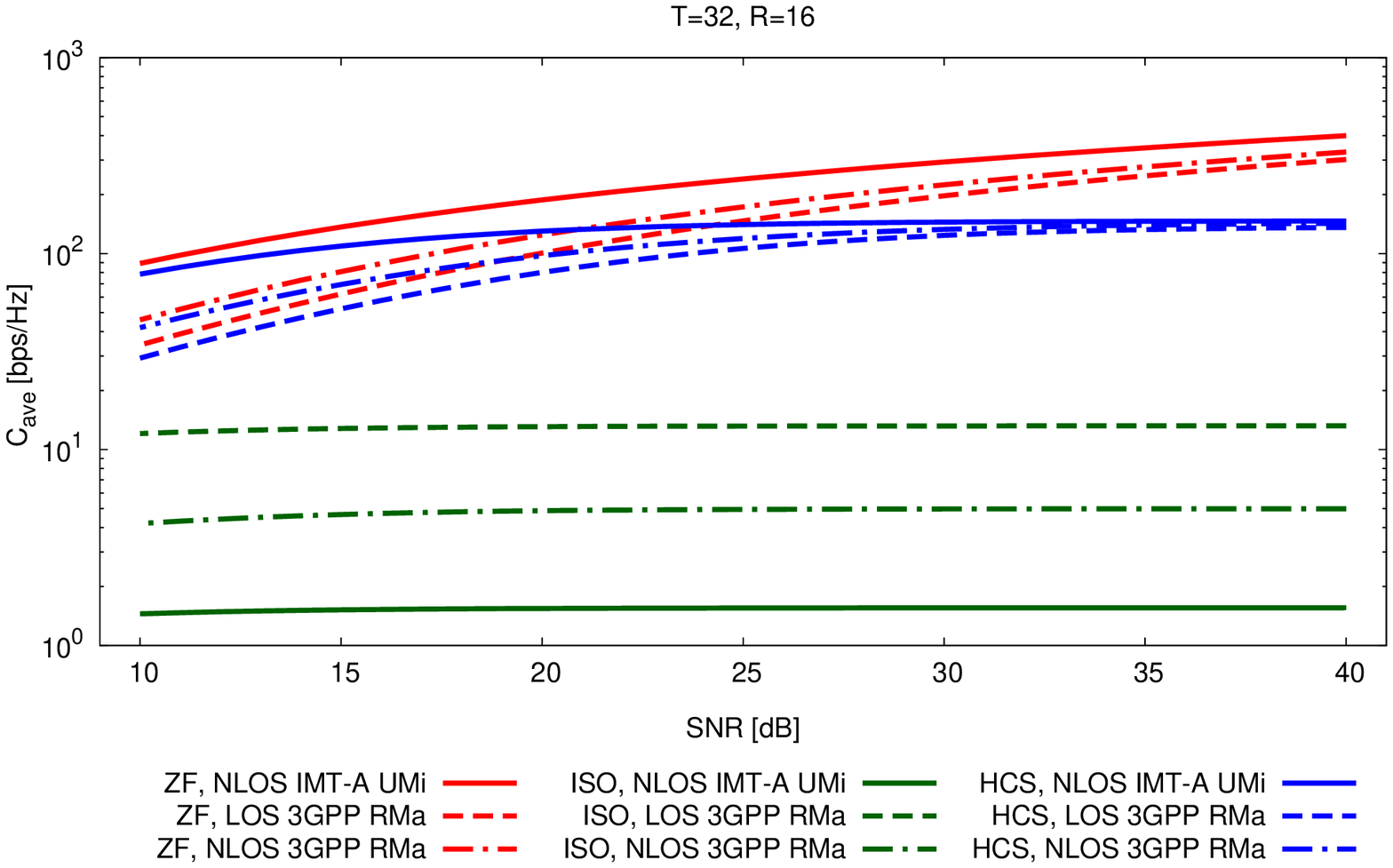}}\tabularnewline
\multicolumn{2}{c}{(\emph{a})}\tabularnewline
\multicolumn{2}{c}{\includegraphics[%
  width=0.95\columnwidth,
  keepaspectratio]{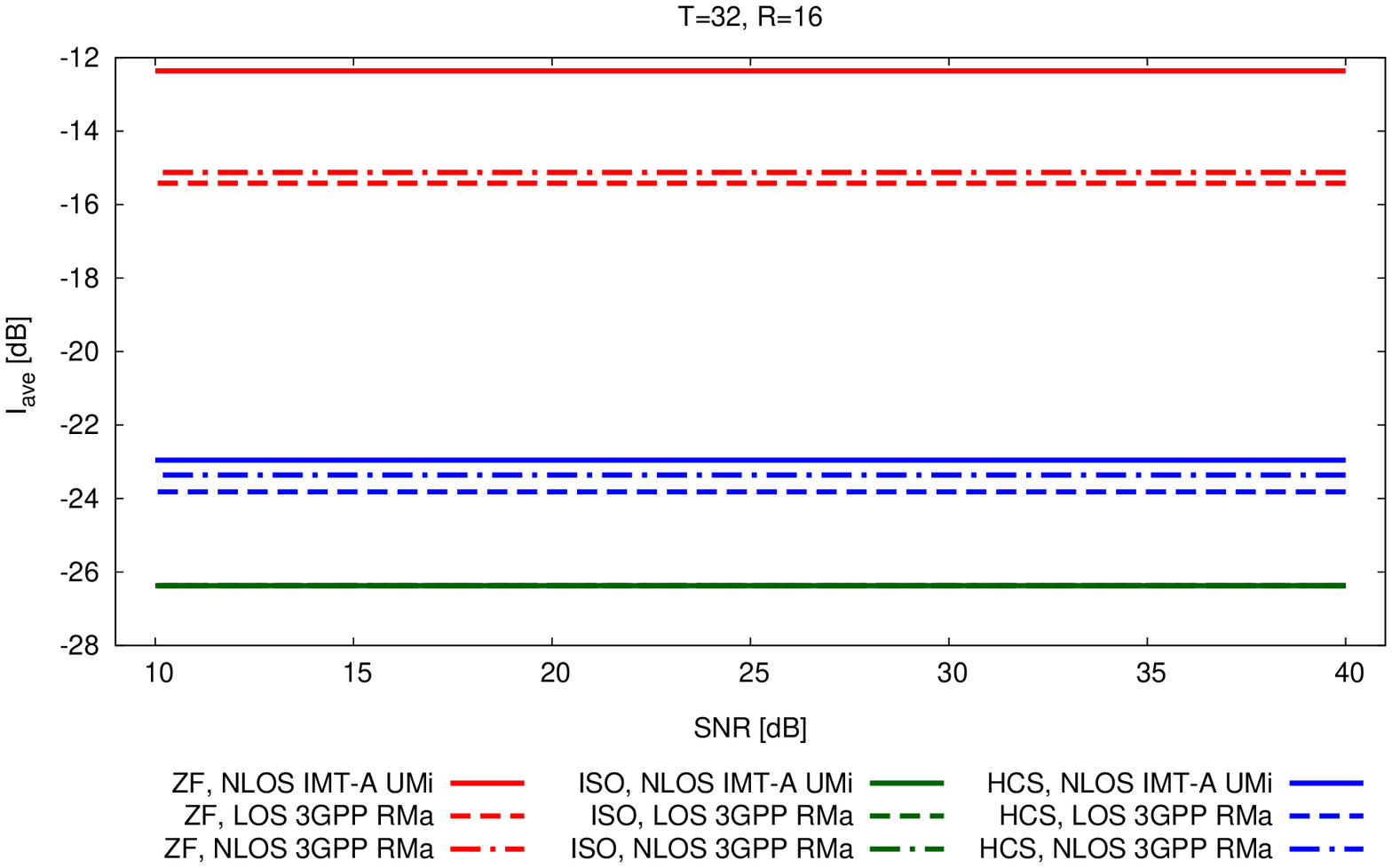}}\tabularnewline
\multicolumn{2}{c}{(\emph{b})}\tabularnewline
\end{tabular}\end{center}

\begin{center}\vfill\end{center}

\begin{center}\textbf{Figure 12 - G. Oliveri et} \textbf{\emph{al.}},
{}``Capacity-Driven Low-Interference Fast Beam Synthesis ...''\end{center}
\newpage

\begin{center}~\vfill\end{center}

\begin{center}\begin{tabular}{cc}
\multicolumn{2}{c}{\includegraphics[%
  width=0.90\columnwidth,
  keepaspectratio]{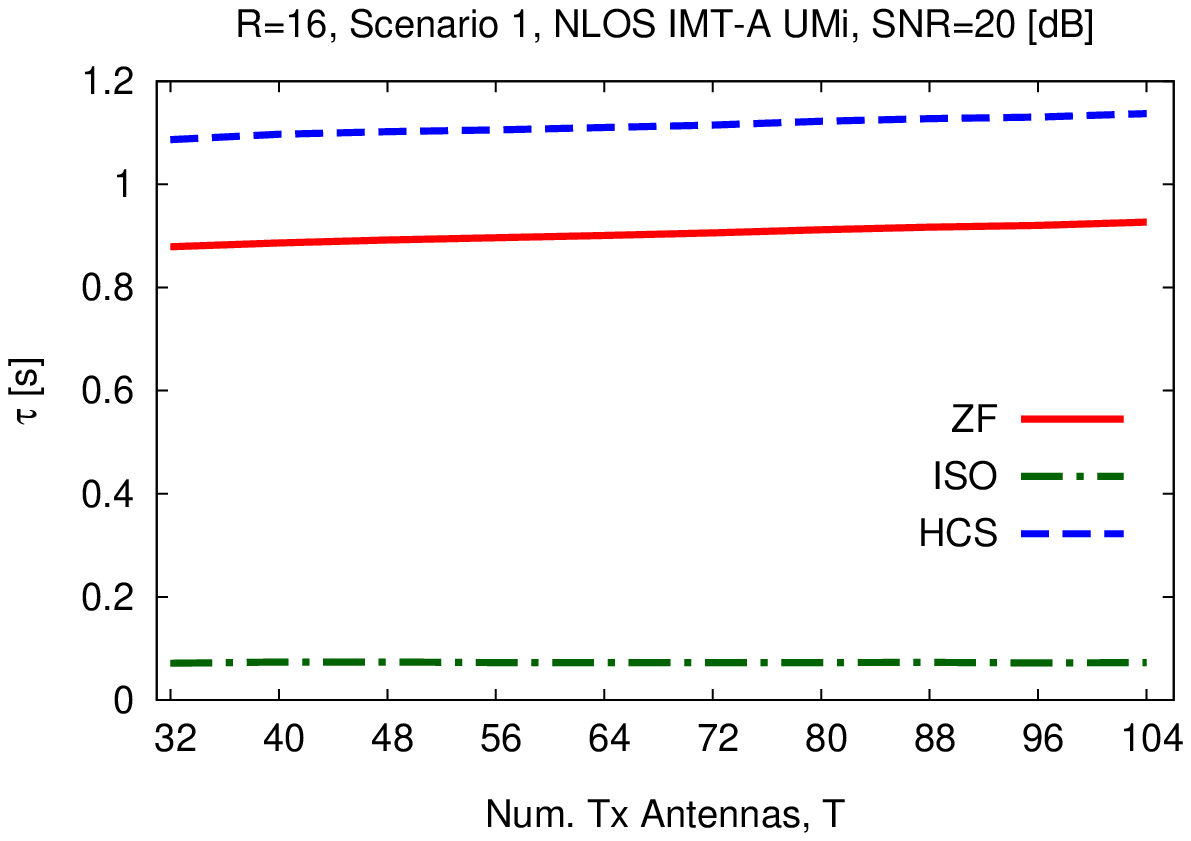}}\tabularnewline
\multicolumn{2}{c}{(\emph{a})}\tabularnewline
\multicolumn{2}{c}{\includegraphics[%
  width=0.90\columnwidth,
  keepaspectratio]{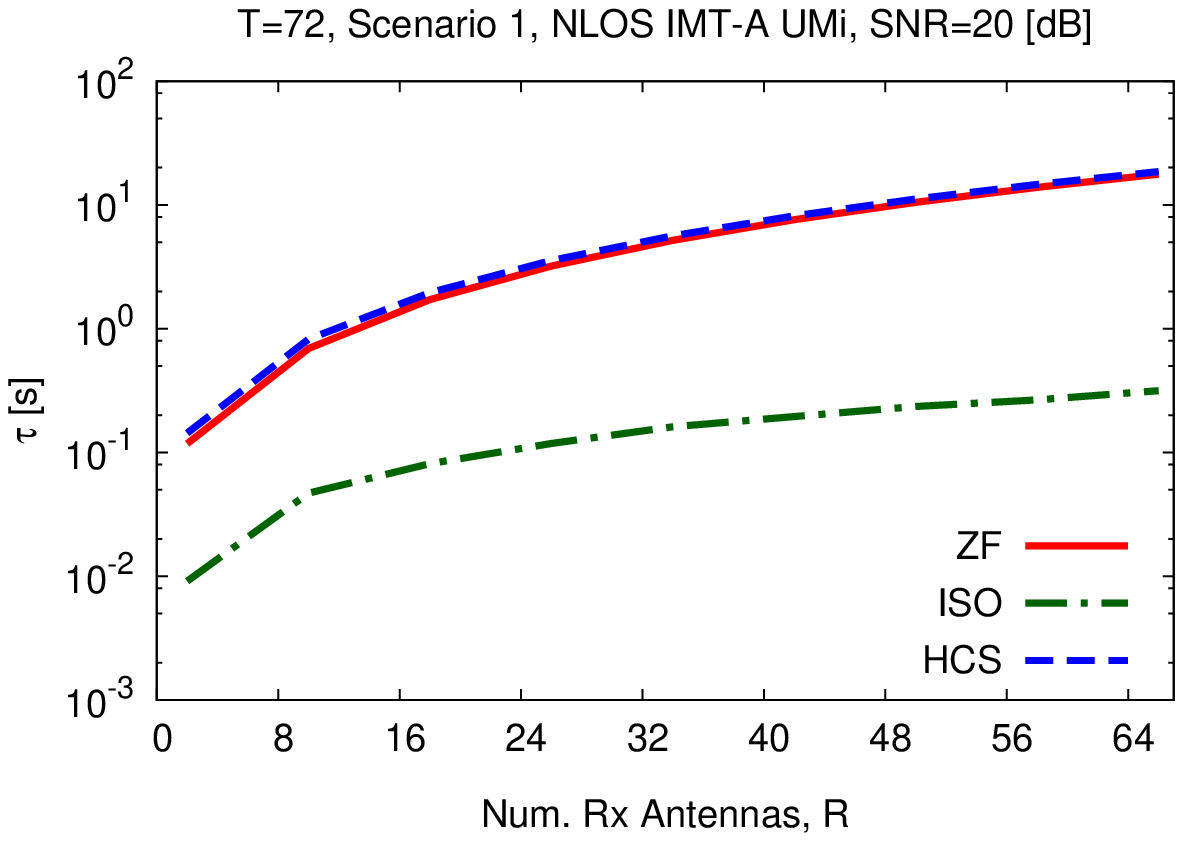}}\tabularnewline
\multicolumn{2}{c}{(\emph{b})}\tabularnewline
\end{tabular}\end{center}

\begin{center}\vfill\end{center}

\begin{center}\textbf{Figure 13 - G. Oliveri et} \textbf{\emph{al.}},
{}``Capacity-Driven Low-Interference Fast Beam Synthesis ...''\end{center}
\newpage

\begin{center}~\vfill\end{center}

\begin{center}\begin{tabular}{|c|c|c|c|}
\hline 
Approach&
$\mathcal{C}_{ave}$ {[}bps/Hz{]}&
$\mathcal{D}_{ave}$ {[}dB{]}&
$\mathcal{I}_{ave}$ {[}dB{]}\tabularnewline
\hline
\hline 
\emph{ZF}&
$187.75$&
$12.35$&
$-12.36$\tabularnewline
\hline 
\emph{ISO}&
$1.55$&
$21.74$&
$-26.37$\tabularnewline
\hline
\emph{HCS}&
$130.09$&
$12.38$&
$-22.95$\tabularnewline
\hline
\end{tabular}\end{center}

\begin{center}\vfill\end{center}

\begin{center}\textbf{Table I - G. Oliveri et} \textbf{\emph{al.}},
{}``Capacity-Driven Low-Interference Fast Beam Synthesis ...\end{center}
\newpage

\begin{center}~\vfill\end{center}

\begin{center}\begin{tabular}{|c|c|c|c|}
\hline 
Approach&
$\mathcal{C}_{ave}$ {[}bps/Hz{]}&
$\mathcal{D}_{ave}$ {[}dB{]}&
$\mathcal{I}_{ave}$ {[}dB{]}\tabularnewline
\hline
\hline 
\emph{ZF}&
$179.16$&
$12.40$&
$-11.73$\tabularnewline
\hline
\emph{ISO}&
$1.46$&
$21.26$&
$-22.34$\tabularnewline
\hline
\emph{HCS}&
$81.37$&
$13.05$&
$-21.82$\tabularnewline
\hline
\end{tabular}\end{center}

\begin{center}\vfill\end{center}

\begin{center}\textbf{Table II - G. Oliveri et} \textbf{\emph{al.}},
{}``Capacity-Driven Low-Interference Fast Beam Synthesis ...\end{center}
\end{document}